\documentclass[a4paper,fleqn,usenatbib,useAMS]{mnras}

\pdfoutput=1
\usepackage{newtxtext,newtxmath}

\usepackage[T1]{fontenc}
\usepackage[applemac]{inputenc}
\usepackage{ae,aecompl}

\usepackage{graphicx} 
\usepackage{amsmath}  
\usepackage{amssymb}  

\def\mean#1{\left< #1 \right>}

\title[Kinematics of lensed galaxies]{A novel 3D technique to study the kinematics of lensed galaxies}

\author[Rizzo et al.]{
Francesca Rizzo,$^{1}$\thanks{E-mail: frizzo@mpa-garching.mpg.de}
Simona Vegetti,$^{1}$
Filippo Fraternali$^{2,3}$
and Enrico Di Teodoro$^{4}$
\\
$^{1}$Max-Planck Institute for Astrophysics, Karl-Schwarzschild Str. 1, D-85748, Garching, Germany\\
$^{2}$University of Groningen, Kapteyn Astronomical Institute, Postbus 800, 9700 AV Groningen, The Netherlands\\
$^{3}$Dipartimento di Fisica e Astronomia, Universita di Bologna, 6/2, Viale Berti Pichat, 40127 Bologna, Italy\\
$^{4}$Research School of Astronomy and Astrophysics-The Australian National University, Canberra, ACT, 2611, Australia\\
}

\pubyear{2018}

\begin{document}
\label{firstpage}
\pagerange{\pageref{firstpage}--\pageref{lastpage}}
\maketitle

\begin{abstract}
We present a 3D Bayesian method to model the kinematics of strongly lensed galaxies from spatially-resolved emission-line observations. This technique enables us to simultaneously recover the lens-mass distribution and the source kinematics directly from the 3D data cube. We have tested this new method with simulated OSIRIS observations for nine star-forming lensed galaxies with different kinematic properties. The simulated rotation curves span a range of shapes which are prototypes of different morphological galaxy types, from dwarf to massive spiral galaxies. We have found that the median relative accuracy on the inferred lens and kinematic parameters are at the level of 1 and 2 per cent, respectively. We have also tested the robustness of the technique against different inclination angles, signal-to-noise ratios, the presence of warps or non-circular motions and we have found that the accuracy stays within a few per cent in most cases. This technique represents a significant step forward with respect to the methods used until now, as the lens parameters and the kinematics of the source are derived from the same 3D data. This enables us to study the possible degeneracies between the two and estimate the uncertainties on all model parameters consistently.  
\end{abstract}

\begin{keywords}
methods: data analysis -- gravitational lensing: strong -- galaxies: kinematics and dynamics -- galaxies: high-redshift
\end{keywords}




\section{Introduction} 
Measuring the content of baryons and dark matter within galaxies, and its evolution with redshift, is a key test of galaxy formation models. In the context of $\Lambda$CDM cosmology, current numerical simulations have not yet produced consistent predictions on the fraction of dark matter within young galaxies. In particular, the amount of dark matter fraction within the stellar half-mass radius has been shown to be strongly dependent on the implementation of feedback processes \citep[e.g.][]{wu, remus, teklu}. For example, different simulations \citep[e.g.][]{lovell, teklu} have resulted in dark matter fraction at z$\sim$2 that can differ by almost one order of magnitude. Numerous physical mechanisms, such as the mass accretion history, the initial mass function, dynamical instabilities, and adiabatic contraction determine the relative contribution of baryons and dark matter within a galaxy \citep[e.g.][]{blue, dutton11, courteau15, zolotov}. For this reason, quantifying the amount of dark matter from kinematical measurements provides a strong constraints on galaxy formation models \citep{rubin, vanalbada}.\\
From an observational perspective, a number of observations have revealed that a significant number of high-redshift galaxies is a disc-dominated system \citep[e.g][]{forster06, forster2009, wisnioski, mason}. However, while in the Local Universe the flatness of the rotation curves and the matter content of disc galaxies is a well-established fact, at high-redshift it is currently a matter of debate, with rotation curves having flat \citep{diteodoro16}, rising \citep{tiley} or declining shapes \citep{lang, genzel17}. The declining rotation curves for six star-forming galaxies at redshift between 0.8 and 2.3 have been explained, for example, by \citet{lang} and \citet{genzel17} as an indication of baryon dominated systems, with a fraction of dark matter lower than 0.2. On the other hand, \citet{diteodoro16}, \citet{diteodoro18}, \citet{mason} have derived rotation curves and velocity dispersion values in agreement with those of local star-forming galaxies \citep{epinat10}.\\
The partition of the matter content between dark matter, stars and gas within a galaxy is provided, also, by studying the evolution of the stellar mass Tully-Fisher relation \citep[TFR,][]{tfr}, which correlates the stellar mass to the rotation velocity, a tracer of the total dynamical mass. A change in the normalisation with redshift might, for example, indicate a redistribution of the total mass between visible and dark matter. \\
Even if the TFR has been explored at redshifts between 0 and 4 by numerous studies, there is no consensus whether it evolves \citep[e.g.][]{puech, straat, ubler, turner2} or not \citep[e.g.][]{miller, miller2, diteodoro16, harrison} with redshift. \\

The diverging results on the kinematics of high-redshift galaxies and, as a consequence of their matter content, can be ascribed to the different methods used to overcome the observational limitations. The study of the kinematics is mainly hampered by two factors: low spatial resolutions and low signal-to-noise ratios (SNR). Seeing-limited observations are typically characterised by an effective spatial resolution of 5 kpc at the redshifts of the sources, z $\sim$ 1-2 \citep[e.g.][]{forster2009, swinbank17}, while a handful of adaptive optics (AO) observations have achieved higher resolutions of $\sim$1-1.6 kpc \citep{molina17, forster18}. Furthermore, because of cosmological surface-brightness dimming, only the bright central regions of galaxies can be observed, especially with AO. Although AO observations are characterized by a better spatial resolution with respect to seeing-limited observations, they have a worse sensitivity and a data binning is often required to increase the SNR.\\
One of the consequences of limited spatial resolution is to smooth out the measured rotation velocity via the so-called beam-smearing effect and can result in an overestimation of the velocity dispersion \citep[e.g.][]{wright, newman, barolo}. This effect can also lead to a misclassification of objects: for example, \citet{newman} have shown that the fraction of dispersion-dominated galaxies in the SINS/zC-SINF surveys \citep{forster2009, cresci09, genzel11} drops from 41 per cent at a seeing-limited resolution to 6-9 per cent when galaxies are observed in the AO mode.\\
The observational limitations imposed by low resolution and signal-to-noise ratio can be successfully overcome by targeting strongly gravitationally lensed galaxies. Strong gravitational lensing offers the opportunity to study high-redshift galaxies at a much higher physical resolution and SNR in their source plane \citep[e.g.][]{nesvadba, swinbank07}. Furthermore, the magnifying power of gravitational lensing opens the possibility to study galaxies in the low-stellar-mass range of $5\times10^8$ - $5\times10^9 M_\odot$ \citep[e.g.][]{jones10, leeth, mason}, which is instead not easily achievable by surveys targeting unlensed galaxies \citep[e.g.][]{forster06, swinbank}. 

It was only in recent years that the potential of gravitational lensing has started to be exploited: for example, \citet{stark08} have studied the kinematics of a lensed galaxy at a resolution of 120 pc at z=3.07. The analysis of two larger samples then followed this study: \citet{jones10} have analysed six lensed galaxies in the redshift range 1.7-3.1, and \citet{livermore} have further extended this sample to 17 targets with redshift from 1 to 4. \citet{leeth} have analysed 15 lensed galaxies at z$\sim$2. Regarding the galaxy population properties,  \citet{jones10} and \citet{leeth} have used different methods to distinguish well-ordered velocity fields from disturbed/merging kinematics and obtained a different classification for similar ranges of redshift and stellar mass: 36 per cent of the galaxies in the \citet{leeth} sample are rotationally dominated and as many as 66 per cent in the \citet{jones10} sample, as confirmed by \citet{livermore}.

So far, most of the analysis aimed at studying the kinematics of lensed sources with optical emission lines, have been characterised by the following features:

\begin{enumerate}
\item the lens mass model is derived from high-spatial-resolution-imaging data \citep[e.g. from HST images,][]{stark08, jones102, jones13, shirazi, leeth, livermore, yuan};
\item the kinematic modelling is done either by delensing the 3D IFU data \citep[e.g.][]{jones13, livermore} and deriving the velocity and dispersion maps with a Gaussian fit to the emission lines in the source plane, or by deriving the moment maps in the image plane and then delensing these maps to the source plane \citep[e.g.][]{jones102, leeth}. In both cases, the lens model is kept fixed.
\item A functional form, usually an arctangent function, is used to fit the delensed velocity field and derive the rotation curve.
\end{enumerate}

Recent studies based on molecular line observations have used a similar approach \citep[e.g.][]{dye, rybak, swinbank15}. One first derives the lens mass distribution from the radio continuum, observed in the same bands as the molecular lines. Then, this model is used to derive the 3D-line data and calculate the corresponding moment maps in the source plane. Finally, kinematic parameters are derived either by applying the kinemetry method \citep{kinemetry} to both the first- and second-moment maps \citep{rybak} or by applying a dynamical model to the first-moment map \citep{dye, swinbank15}.

All these approaches are suboptimal mainly for two reasons: first, if the lens model is kept fixed it is not possible to quantify any degeneracy between the lens mass parameters and the source kinematic properties. Second, the kinematic fitting is done on the reconstructed source rather than on the data. However, on the source plane, pixels are correlated, the noise properties not fully characterised and the effective resolution changes with position according to the lensing magnification. As a consequence, one introduces systematic errors in the derivation of the kinematic properties of the source, which may be difficult to quantify.

Recently, \citet{patricio} have applied a forward modelling approach which partly overcomes some of the above issues by deriving the velocity map directly on the image plane through a Gaussian fitting to the emission lines. However, similarly to the techniques described above, this method is not ideal, as it relies on a fixed lens model derived from a separate {\it HST} observation and it performs a kinematic modelling of the 2D velocity map, instead of the full 3D data cube. 

Finally, other studies have been focusing on sources that are not significantly lensed (i.e. only weakly distorted), so that the kinematic analysis can be done directly on the image plane without having to reconstruct the unlensed emission \citep[e.g.][]{mason, girard, diteodoro18}. However, even small distortions of the observed axis ratio, due to lensing, could affect the capability to recover the kinematic parameters accurately. \citet{mason} have tried to correct for this effect using a global value for the magnification factor.

In this paper, we present a novel Bayesian three-dimensional and pixellated approach, which, applied either to IFU or interferometric data, enables us to simultaneously reconstruct both the lensing mass distribution and the kinematics of the source. Our method represents a significant improvement over those described above as it is not affected by differential magnification nor poor understanding of the errors and resolution properties of the reconstructed unlensed plane. Our technique does not require the use of high-resolution imaging data for the derivation of the lens parameters, as these are derived from the same 3D data used for the modelling of the kinematics of the background galaxy. Since the lens parameters and the source are inferred simultaneously from the same dataset, our method is not affected by differential magnification. Moreover, the kinematics of the background galaxy is not obtained by fitting on the source plane, but directly the lensed data in a hierarchical Bayesian fashion, where the kinematics on the source plane is essentially an hyper-parameter (i.e., parameter defining the prior) of the model. The main novelty of our procedure is that a modified tilted-ring kinematic model is an extra constraint for a pixellated source reconstruction. Furthermore, the derivation of the lens mass model and the source kinematics is done simultaneously, allowing us to quantify possible degeneracies and to estimate the errors on all model parameters using a Bayesian approach. Finally, our 3D approach enables us to describe the kinematics of the source minimazing the influence of the beam smearing effect.

This paper is organised as follows. In Section \ref{sec:method}, we describe in details the method used for the lens modelling and the derivation of the kinematics. In Section \ref{sec:mock}, we present the IFU simulated datasets. In Section \ref{sec:modelling} we describe the modelling strategy and the assumptions applied to model the simulated datasets, which are then used in Section \ref{sec:results} to test our method under different observational set-ups. The robustness and the limits of the technique are summarised in Section \ref{sec:conclusions}, where we also list future developments and applications.

\section{Method description}
\label{sec:method}

This section describes the core features of our method, which is an extension  of the technique developed by \citet{vegetti09} to the 3D-domain.
In particular, we present the statistical framework that allows us to reconstruct the background source, its kinematics and the lens mass distribution. 
The lens-mass parametrisation is described in Section \ref{sec:lens}, while the details of the kinematic model used to describe the lensed source are given in Section \ref{sec:kinematic_model}.

\subsection{Source reconstruction}
\label{sec:sourcerec}
\begin{figure*}
\includegraphics[width=2\columnwidth]{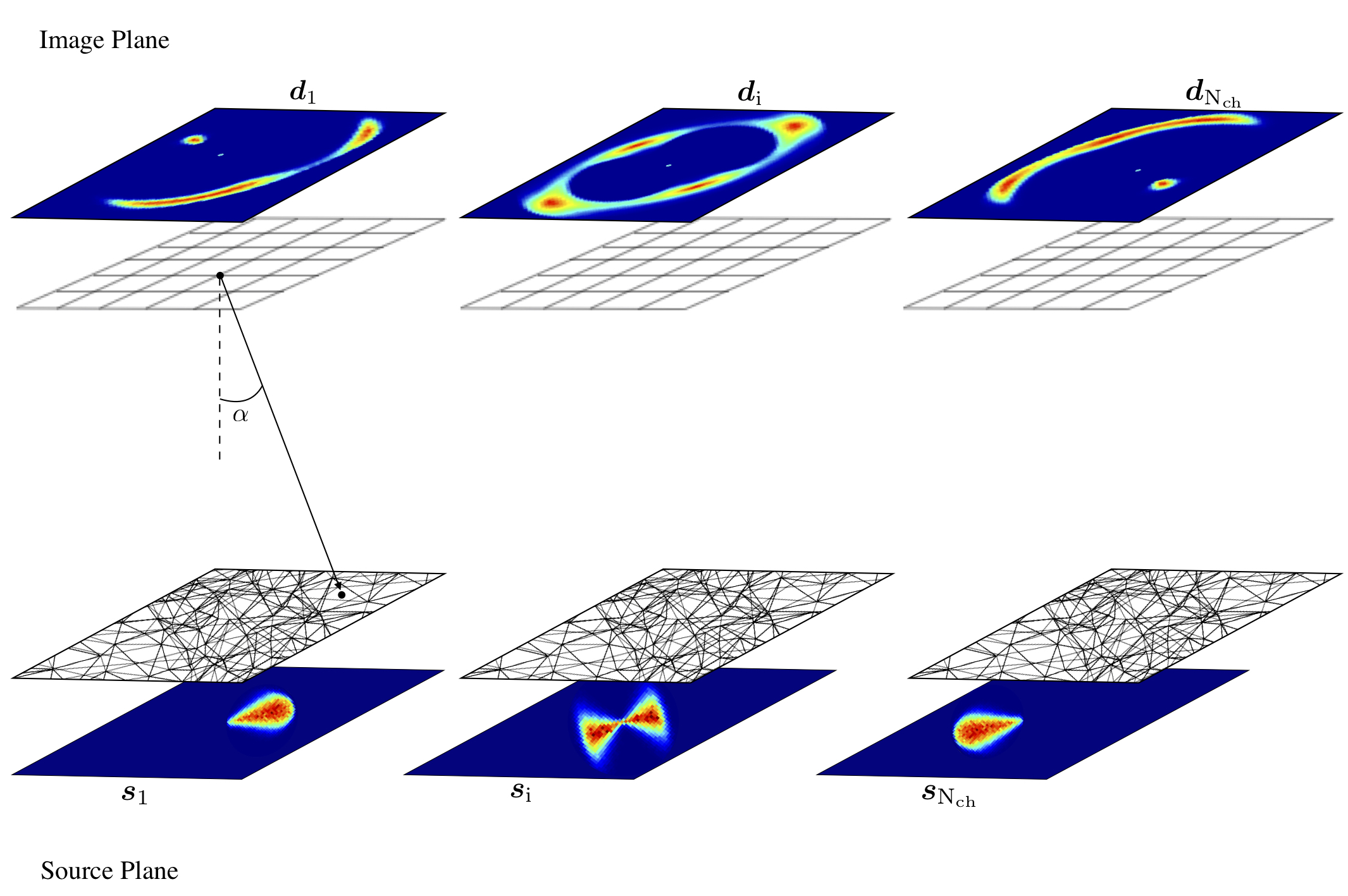}
\caption{A schematic view of the source and lens planes. On the upper panel, the lensed data for three representative spectral channels and the respective regular grid on the image plane. For each spectral channel, the position $\vec{x}$ of a subset of $\mathrm{N_{s}}$ pixels in the image plane corresponds to a position $\vec{y}$ on the source plane (lower panel), through the lens equation $\vec{y}=\vec{x}-\vec{\alpha}\left(\vec{x}\right)$. The points $\vec{y}$ are the vertices of a triangular adaptive grid on the source plane.}
\label{fig:lens}
\end{figure*}

In the following, we indicate with $\mathbf{s}$ and $\mathbf{d}$ the 3D pixellated surface brightness distribution of the source and the data in the image plane, respectively. We refer the reader to Figure \ref{fig:lens} for a schematic representation of the source and image planes. 

Given a set of $\mathrm{N_{ch}}$ observed spectral channels, the source and data vectors, $\mathbf{s}$ and $\mathbf{d}$, have a total of $\mathrm{N_{ch}}$ components, $\mathbfit{s}_{\mathrm{i}..\mathrm{N_{ch}}}$ and $\mathbfit{d}_{\mathrm{i}..\mathrm{N_{ch}}}$ respectively, each representing the surface brightness distribution in one channel $\mathrm{i}$: 
\begin{equation}
\mathbf{s}=\{ \mathbfit{s}_{1}, ..., \mathbfit{s}_{\mathrm{i}}, ..., \mathbfit{s}_\mathrm{N_{ch}}\},
\end{equation}
\begin{equation}
\mathbf{d}=\{ \mathbfit{d}_{1}, ..., \mathbfit{d}_{\mathrm{i}}, ..., \mathbfit{d}_{\mathrm{N_{ch}}}\}.
\end{equation}
For each -ith spectral channel, the surface brightness distribution of the lensed cube $\mathbfit{d}_{\mathrm{i}} $, its noise $\mathbfit{n}_{\mathrm{i}}$ and the relative unknown source surface brightness distribution $\mathbfit{s}_{\mathrm{i}}$ are related to each other by the following set of linear equation:
\begin{equation}
\mathbfss{M}\left(\boldsymbol{\psi}\right)\mathbfit{s}_{\mathrm{i}}+ \mathbfit{n}_{\mathrm{i}}=\mathbfit{d}_{\mathrm{i}},
\label{eq:lens}
\end{equation}
where $\mathbfss{M=B\,L}$ is the response operator which depends on the lensing operator $\mathbfss{L}$ and the point spread function (PSF) operator $\mathbfss{B}$ (for interferometric data, $\mathbfss{B}$ is the Fourier transfer operator). The lensing operator $\mathbfss{L}$, is a non-linear function of the lens mass-density distribution parameters $\boldsymbol{\eta_{\mathrm{lens}}}$ (see Section \ref{sec:lens} for their definition) via the lensing potential $\psi\left(\boldsymbol{\eta_{\mathrm{lens}}}\right)$.\\
The method used for the source reconstruction is grid-based in the sense that the background-source surface-brightness-distribution is reconstructed on a triangular adaptive grid defined by a Delaunay tessellation. The source grid automatically adapts with the lensing magnification, so that there is a high pixel-density in the high-magnification regions close to the caustics. The vertices of the triangular grid are obtained by casting back to the source plane a subset $\mathrm{N_{s}}$ of the $\mathrm{N_{d}}$ image-plane pixels via the lens equation. We determine the surface brightness at each source-plane pixel by interpolating between the values at the vertices of the triangles. We reconstruct each channel on the same triangulation.

As both $\boldsymbol{\eta_{\mathrm{lens}}}$ and the source $\mathbfit{s}_{\mathrm{i}}$ are unknown, equation (\ref{eq:lens}) is ill-defined and cannot be simply inverted. Therefore, we derive a penalty function defined in the context of three levels of Bayesian inference, which are described below.

\subsubsection{First level of inference - Linear optimization}

Using Bayes' rule, the most probable a posteriori source, $\mathbf{s_{\rm MP}}$, given the data and a lens mass model, is derived by maximising the following posterior probability \citep{ suyu,vegetti09}:
\begin{equation}
P\left(\mathbf{s}|\mathbf{d}, \lambda, \boldsymbol{\eta_{\mathrm{lens}}}, \boldsymbol{\eta_{\mathrm{kin}}},\mathbfss{R}\right)=\frac{P\left(\mathbf{d}|\mathbf{s}, \boldsymbol{\eta_{\mathrm{lens}}}\right) P\left(\mathbf{s} | \lambda, \boldsymbol{\eta_{\mathrm{kin}}}, \mathbfss{R}\right)}{P \left(\mathbf{d}| \lambda, \boldsymbol{\eta_{\mathrm{lens}}}, \boldsymbol{\eta_{\mathrm{kin}}}, \mathbfss{R}\right)}.
\label{eq:posterior}
\end{equation}
Here, the matrix $\mathbfss{R}$ is the source regularization form (variance, gradient or curvature), with a strength set by the regularization level vector $\lambda$ \citep[see][for further details]{koopmans,vegetti09}. The regularization level vector $\lambda$ has $\mathrm{N_{ch}}$ components, so that the user can choose whether the level of regularisation is constant across the spectral channels or not. For simplicity, all equations below and above assume a constant value of $\lambda$ as a function of frequency. $\boldsymbol{\eta_{\mathrm{kin}}}$ in equation (\ref{eq:posterior}) are the source kinematic parameters, as defined in Section \ref{sec:kinematic_model}. The remaining terms in equation (\ref{eq:posterior}) are the likelihood function $P\left(\mathbf{d}|\mathbf{s}, \boldsymbol{\eta_{\mathrm{lens}}}\right)$, the prior on the source surface brightness distribution $P\left(\mathbf{s} | \lambda, \boldsymbol{\eta_{\mathrm{kin}}}, \mathbf{R}\right)$ and the evidence $P\left(\mathbf{d}| \lambda, \boldsymbol{\eta_{\mathrm{lens}}},\boldsymbol{\eta_{\mathrm{kin}}}, \mathbfss{R}\right)$, which is irrelevant for the maximization of the posterior, but plays an important role at the third level of inference (see Section \ref{sec:optimisation}). Under the the assumption of Gaussian noise, the likelihood can be expressed as follows
\begin{equation}
P\left(\mathbf{d}|\mathbf{s}, \boldsymbol{\eta_{\mathrm{lens}}}\right)=\frac{\exp\left[-E_{\mathrm{D}}\left(\mathbf{d}|\mathbf{s}, \boldsymbol{\eta_{\mathrm{lens}}}\right)\right]}{Z_{\mathrm{D}}},
\label{eq:like}
\end{equation}
where $Z_{\mathrm{D}}$ is the normalization, and $E_{\mathrm{D}}$ is half the standard $\chi^2$ and is given by 
\begin{equation}
E_{\mathrm{D}}\left(\mathbf{d}|\mathbf{s}, \boldsymbol{\eta_{\mathrm{lens}}}\right)=\frac{1}{2} \sum_{\mathrm{i}=1}^{\mathrm{Nch}}\left(\mathbfss{M}\mathbfit{s}_{\mathrm{i}}-\mathbfit{d}_{\mathrm{i}}\right)^{\intercal}\mathbfss{C}_{\mathrm{d\,i}}^{-1}\left(\mathbfss{M}\mathbfit{s}_{\mathrm{i}}-\mathbfit{d}_{\mathrm{i}}\right).
\label{eq:ed}
\end{equation}
Above, $\mathbfss{C}_{\mathrm{d\,i}}$ is the covariance matrix of the data for the -ith spectral channel. Since the noise is assumed to be uncorrelated, it is a diagonal matrix.\\
In our implementation, the source prior assumes a quadratic form, which peaks at a source kinematic model, $\mathbf{s_{\mathrm{kin}}\left(\boldsymbol{\eta_{\mathrm{kin}}}\right)}$. The prior probability distribution is, therefore, expressed as
\begin{equation}
P\left(\mathbf{s} | \lambda, \boldsymbol{\eta_{\mathrm{kin}}}, \mathbfss{R}\right)=\frac{\exp\left[-\lambda E_{\mathrm{R}}\left(\mathbf{s}|\boldsymbol{\eta_{\mathrm{kin}}}, \mathbfss{R}\right)\right]}{Z_{\mathrm{R}}},
\label{eq:prior}
\end{equation}
where the quadratic functional $E_{\mathrm{R}}$ and the normalization $Z_{\mathrm{R}}$ are given respectively by
\begin{equation}
E_{\mathrm{R}}\left(\mathbf{s}\right)= \sum_{\mathrm{i}=1}^{\mathrm{Nch}} \left[E_{\mathrm{R}}\left(\mathbfit{s}_{\mathrm{kin\,i}}\right)+\frac{1}{2}\left(\mathbfit{s}_{\mathrm{i}}-\mathbfit{s}_{\mathrm{kin\,i}}\right)^{\intercal} \mathbfss{H}_{\mathrm{R, i}}\left(\mathbfit{s}_{\mathrm{i}}-\mathbfit{s}_{\mathrm{kin\,i}}\right)\right]
\label{eq:es}
\end{equation}
and   
\begin{equation}
Z_{\mathrm{R}}\left(\lambda\right)=\int d^{N_{s}}\mathbf{s}\,e^{-E_{\mathrm{R}}\left(\mathbf{s}\right)}=e^{-E_{\mathrm{R}}\left(\mathbf{s_{kin}}\right)}\left(\frac{2\pi}{\lambda} \right)^{\frac{\mathrm{N_{s}N_{ch}}}{2}}\left(\det \mathbfss{H}_{\mathrm{R}}\right)^{-\mathrm{N_{ch}/2}}.
\label{eq:zs}
\end{equation}
$\mathbfss{H}_{\mathrm{R}}$ in the above equation is a block matrix made up of the N$_{\mathrm{ch}}$ matrices $\mathbfss{H}_{\mathrm{R, i}}$ (see equation~\ref{eq:es}) and it is defined as the Hessian of $E_{\mathrm{R}}$, $\mathbfss{H}_{\mathrm{R}}=\nabla \nabla E_{\mathrm{R}}=\mathbfss{R}^{\intercal}\mathbfss{R}$. The most probable surface brightness is obtained by maximising the posterior probability in equation~(\ref{eq:posterior}), i.e. by solving the following set of linear equations:
\begin{equation}
\left[\mathbfss{M}^{\intercal}\mathbfss{C}_{\mathrm{d\,i}}^{-1}\mathbfss{M}+\lambda \mathbfss{H}_{\mathrm{R}}\right]\mathbfit{s}_{i}=\mathbfss{M}^{\intercal}\mathbfss{C}_{\mathrm{d\,i}}^{-1}\mathbfit{d}_{\mathrm{i}}+\lambda \mathbfss{H}_{\mathrm{R}}\mathbfit{s}_{\mathrm{kin\,i}}.
\label{eq:lin_k}
\end{equation}
The form of these equations differs from those derived in \citet{vegetti09} in the last term on the right-hand side, $\lambda \mathbfss{H}_{\mathrm{R i}}\mathbfit{s}_{\mathrm{kin\,i}}$. This term is due to a different assumption about the peak of the source prior, which is equal to zero in \citet{vegetti09} and $\mathbf{s_{\mathrm{kin}}}$ here.

\subsubsection{Second level of inference - Non-linear optimization}

To infer the kinematic parameters $\boldsymbol{\eta_{\mathrm{kin}}}$, the lens parameters $\boldsymbol{\eta_{\mathrm{lens}}}$ and the optimal level of regularization $\lambda$ we maximize the following posterior probability 
\begin{equation}
P\left(\lambda, \boldsymbol{\eta_{\mathrm{kin}}}, \boldsymbol{\eta_{\mathrm{lens}}}|\mathbf{d},\mathbfss{R}\right)=\frac{P\left(\mathbf{d}|\lambda, \boldsymbol{\eta_{\mathrm{lens}}}, \boldsymbol{\eta_{\mathrm{kin}}}, \mathbfss{R}\right) P\left(\lambda, \boldsymbol{\eta_{\mathrm{kin}}}, \boldsymbol{\eta_{\mathrm{lens}}}\right)}{P\left(\mathbf{d}|\mathbfss{R}\right)}.
\label{eq:posterior_nl}
\end{equation}
Assuming a prior which is flat in log$\lambda$, $\boldsymbol{\eta_{\mathrm{lens}}}$ and $\boldsymbol{\eta_{\mathrm{kin}}}$, equation~(\ref{eq:posterior_nl}) can be expressed as 
\begin{equation}
P\left(\mathbf{d}|\lambda, \boldsymbol{\eta_{\mathrm{lens}}},\boldsymbol{\eta_{\mathrm{kin}}}, \mathbf{R}\right)= \int  \, P\left(\mathbf{d}|\mathbf{s}, \boldsymbol{\eta_{\mathrm{lens}}}\right)P\left(\mathbf{s} | \lambda, \boldsymbol{\eta_{\mathrm{kin}}}, \mathbfss{R}\right)\, d\mathbf{s}.
\label{eq:post_int}
\end{equation}
If we assume a Gaussian noise and a quadratic form of regularization, equation~(\ref{eq:post_int}) can be written as 
\begin{equation}
\begin{split}
P\left(\mathbf{d}|\lambda, \boldsymbol{\eta_{\mathrm{lens}}}, \boldsymbol{\eta_{\mathrm{kin}}}, \mathbf{R}\right)=-E\left(\mathbf{s_{MP}}\right)-\frac{\mathrm{N_{ch}}}{2}\log{\det \mathbfss{H}_{\mathrm{E}}} +\frac{\mathrm{N_{s}N_{ch}}}{2}\log{\lambda}\\ +\lambda E_{\mathrm{s}}\left(\mathbf{s_{kin}}\right)+\frac{\mathrm{N_{ch}}}{2}\log{\det \mathbfss{H}_{\mathrm{R}}}-\frac{\mathrm{N_{d}N_{ch}}}{2}\log{2\pi}-\frac{1}{2}\sum_{\mathrm{i}=1}^{\mathrm{Nch}} \log{\det\mathbfss{C}_{\mathrm{d\,i}}}.
\end{split}
\label{eq:p_lam}
\end{equation}
In the above equation, $E=E_{\mathrm{D}}+\lambda E_{\mathrm{R}}$, $\mathbfss{H}_{\mathrm{E}}$ is its Hessian and $\mathbf{s_{MP}}$ is the most probable solution that maximizes the posterior, $\nabla E \left(\mathbf{s_{MP}}\right) = 0$.\\

The expression for the posterior probability, equation (\ref{eq:p_lam}), differs from that derived by \citet{suyu} and \citet{vegetti09} for the multiplications/summation by/over $\mathrm{N_{ch}}$ and the presence of the term $E_{\mathrm{s}}\left(\mathbf{s_{kin}}\right)$, which is the main novelty of our approach. This allows us to derive the kinematic parameters of the source, while retaining the flexibility of a pixellated source surface brightness distribution, simultaneously infer the lens mass distribution and take advantage of the extra constraints provided by the velocity channels.

\subsubsection{Third level of inference - Model comparison}
\label{sec:third}
At the third level of inference, to compare and rank different models, we calculate the marginalized Bayesian evidence which is a measure of the probability of the data given the model. In our case, this marginalized evidence can be expressed as the integral of the normalization factor in equation~(\ref{eq:posterior}) over the lens parameters $\boldsymbol{\eta_{\mathrm{lens}}}$, the kinematic parameters $\boldsymbol{\eta_{\mathrm{kin}}}$ and the source regularization $\lambda$, such that
\begin{multline}
P\left(\mathbf{d}|\mathbfss{R}\right)= \int P\left(\mathbf{d}|\lambda, \boldsymbol{\eta_{\mathrm{lens}}}, \boldsymbol{\eta_{\mathrm{kin}}}, \mathbfss{R}\right)\times\\
P\left(\lambda, \boldsymbol{\eta_{\mathrm{lens}}},\boldsymbol{\eta_{\mathrm{kin}}}\right)\,d\lambda\,d\boldsymbol{\eta_{\mathrm{lens}}}\,d\boldsymbol{\eta_{\mathrm{kin}}}.
\label{eq:integ}
\end{multline}
This integral is calculated numerically with {\sc MultiNest} \citep{feroz09, feroz13}, which is a Nested Sampling-based method improving on the original idea by \citet{skilling}. As a by-product of this evidence calculation, we also obtain the posterior distributions of the model parameters, allowing us to estimate their statistical uncertainties and degeneracies (see Section \ref{sec:results}).

\subsection{Lens mass model}
\label{sec:lens}

The lens parameters $\boldsymbol{\eta_{\mathrm{lens}}}$ which define the lensing operator $\mathbfss{L}$ are: $\kappa_0$, $q$, $\gamma$, $x_0$, $y_0$, $\theta$, $\Gamma_{\mathrm{sh}}$, $\theta_{\mathrm{sh}}$. These parameters describe a projected mass density profile as a cored elliptical power-law distribution with the contribution of an external shear component of strength $\Gamma_{\mathrm{sh}}$ and position angle $\theta_{\mathrm{sh}}$. The dimensionless projected mass density profile is defined as
\begin{equation}
\kappa\left(x,y\right)=\frac{\kappa_0\left(2-\frac{\gamma}{2}\right)q^{\gamma-\frac{3}{2}}}{2\left[q^2\left(x^2+r_{\mathrm{c}}^2\right)+y^2\right]^{\frac{\gamma-1}{2}}}\,.
\end{equation}
$\kappa_0$ is the surface mass-density normalization, $q$ is the projected flattening, $\gamma$ is the density slope, $x_0$ and $y_0$ define the centre of the mass distribution, $r_{\mathrm{c}}$ is the core radius and $\theta$ is the position angle of the major axis. The Einstein radius for this density profile is defined as
\begin{equation}
R_{\mathrm{ein}}=\left[\frac{\kappa_0\left(2-\frac{\gamma}{2}\right)q^{\frac{\gamma-2}{2}}}{3-\gamma}\right]^{\frac{1}{\gamma-1}}.
\end{equation}
In the following sections, we assume that the mass distribution has a negligible core radius of $10^{-4}$ arcsec.

\subsection{Source kinematic model}
\label{sec:kinematic_model}

We build the kinematic model using a modified version of the building-model function of $^{\mathrm{3D}}$BAROLO \citep{barolo}. To simulate the gas emission from a rotating galaxy the $^{\mathrm{3D}}$BAROLO algorithm uses a stochastic function that populates a 6D domain (three spatial and three spectral dimensions) with emitting gas clouds, that allow us to build the line profiles. The rotating galaxy is modelled as a series of concentric circular rings using the so-called tilted-ring model \citep{rogstad}. On each ring, the position of the clouds are chosen randomly in such a way that on average the clouds become uniformly distributed over its surface. Each ring is described by the following parameters: 
\begin{enumerate}
\item the coordinates of the centre $x_{\mathrm{s}}, y_{\mathrm{s}}$;
\item the inclination angle $i$, defined such that $i=90^{\circ}$ for an edge-on galaxy and $i=0^{\circ}$ for a face-on one; 
\item the position angle $PA$, defined as the angle between the north direction of the sky and the projected major axis of the receding half of the rings measured counterclockwise; 
\item the face-on gas column density $\Sigma$; 
\item the systemic velocity $V_{\mathrm{sys}}$; 
\item the rotation velocity $V_{\mathrm{rot}}$;
\item the velocity dispersion $\sigma_{\mathrm{gas}}$. 
\end{enumerate}
The projected velocity along the line of sight $V_{\mathrm{los}}$ at a particular radius $R$ is defined by
\begin{equation}
V_{\mathrm{los}}\left(R\right)=V_{\mathrm{sys}} +V_{\mathrm{rot}}\left(R\right) \cos{\phi} \sin{i} \,,
\label{eq:vlos}
\end{equation}
where $\phi$ is the azimuthal angle in the plane of the galaxy. 
To build the 3D model, at each radius, the positions of the clouds are then rotated and projected into the plane of the sky with an orientation with respect to the observer defined by both the position and the inclination angles at that radius. As in $^{\mathrm{3D}}$BAROLO, to obtain the velocity profile at each location, the clouds are divided into sub-clouds. Each of these sub-clouds has a velocity which is drawn from a Gaussian distribution with a dispersion of $\sigma^2= \sigma^2_{\mathrm{gas}}+\sigma^2_{\mathrm{instr}}$. Here, $\sigma_{\mathrm{instr}}$ takes into account the instrumental broadening.

Unlike $^{\mathrm{3D}}$BAROLO, our implementation does not allow for a variation of all the parameters ring by ring. Instead, we make the following assumptions: (i) all the rings have the same centre coordinates and systemic velocity (in Section \ref{sec:further_tests} we explicitly test the effects of this assumption); (ii) the radial variation of the inclination and position angles are described by a polynomial of degree from 0 to 3; (iii) the radial variation of the rotation velocity and velocity dispersion are described by functional forms. The use of functional forms for the rotation velocity and velocity dispersion allows us to reduce the number of free parameters. Our kinematic model $\mathbf{s_{\mathrm{kin}}}$ is, therefore, defined by the following set of parameters $\boldsymbol{\eta_{\mathrm{kin}}}=\{R_{\mathrm{ext}}, \Sigma, x_{\mathrm{s}}, y_{\mathrm{s}}, V_{\mathrm{sys}}, i, PA, V_{\mathrm{rot}}, \sigma_{\mathrm{gas}}\}$. $R_{\mathrm{ext}}$ is the radial extension and is a fixed parameter chosen by the user. In Section \ref{sec:modelling} we describe the assumptions made to estimate $R_{\mathrm{ext}}$ for the simulated data analysed in this paper. 
Following $^{\mathrm{3D}}$BAROLO, the surface density of the gas $\Sigma$ is also not a free parameter; instead, we impose a pixel-by-pixel normalisation, which is given by the surface brightness distribution obtained from the lens modelling of the zeroth-moment map. The advantage of using a spatially-changing normalisation is that it allows us to take into account for possible asymmetries in the ionised or molecular gas distribution, as it is typical for high-redshift galaxies given the presence of massive star-forming regions \citep[e.g.][]{genzel11, swinbank12, livermore}. Therefore, the presence of clumpy emissions or holes should not affect the derived kinematics, because this normalization ensures that the kinematics is independent of the gas distribution \citep[e.g.][]{barolo, lelli12}.

By construction, the kinematic model is built on a Cartesian grid, defined by a pixel scale and dimensions chosen by the user. Since the source reconstruction is determined on a Delaunay tessellation (see Section \ref{sec:sourcerec}), we then map this model at the positions of the triangle vertices.

\subsubsection{Rotation velocity and velocity dispersion curves}

We have implemented three empirical functions to describe the rotation curves: the arctangent function \citep{courteau}, the hyperbolic tangent function \citep{tanh} and a multi-parameter function \citep{rix}. These are expressed by the following three expressions, respectively:
\begin{equation}
V_{\mathrm{rot}}\left(R\right) = \frac{2}{\pi}\,V_{\mathrm{t}}\,\arctan\left(\frac{R}{R_{\mathrm{t}}}\right),
\label{eq:acrtg}
\end{equation}
\begin{equation}
V_{\mathrm{rot}}\left(R\right) = V_{\mathrm{t}}\,\tanh\left(\frac{R}{R_{\mathrm{t}}}\right),
\label{eq:hacrtg}
\end{equation}
\begin{equation}
V_{\mathrm{rot}}\left(R\right) = V_{\mathrm{t}}\frac{\left(1+\frac{R_{\mathrm{t}}}{R}\right)^{\beta}}{\left[1+\left(\frac{R_{\mathrm{t}}}{R}\right)^{\xi}\right]^{1/\xi}}.
\label{eq:mult}
\end{equation}
$R_{\mathrm{t}}$ is the turnover radius between the inner rising and outer part of the curve. $V_{\mathrm{t}}$ is the asymptotic velocity for the arctangent and hyperbolic tangent functions, and the velocity scale for the multi-parameter one. $\xi$ defines the sharpness of the turnover while $\beta$ specifies the power-law behaviour of the curve at large radii. The arctangent function has mainly been used to model the kinematics of high-redshift galaxies \citep[e.g.][]{swinbank, leeth}. However, it is not flexible enough to reproduce the different kinds of observed rotation curves, especially in the inner regions where a bump can be present. For this reason, we prefer the multi-parameter function which is, by definition, more flexible. 

To describe the velocity dispersion profile, the user can choose from a power-law, a linear or an exponential function:
\begin{equation}
\sigma_{\mathrm{gas}}\left(R\right) = \sigma_{0}\,\left(\frac{R}{R_{\sigma}}\right)^{\zeta},
\label{eq:power}
\end{equation}
\begin{equation}
\sigma_{\mathrm{gas}}\left(R\right) = \sigma_{0}+\zeta\,R,
\label{eq:linear}
\end{equation}
\begin{equation}
\sigma_{\mathrm{gas}}\left(R\right) = \sigma_{0}\,e^{-\frac{R}{R_{\sigma}}}+\sigma_{1}.
\label{eq:exp}
\end{equation}

\subsection{Optimisation scheme}
\label{sec:optimisation}

We infer the unknown parameters $\boldsymbol{\eta_{\mathrm{lens}}}$, $\lambda$, $\boldsymbol{\eta_{\mathrm{kin}}}$ and the source $\mathbf{s}$ with an optimisation scheme, which is divided in the following four stages (see also Figure \ref{fig:flowchart} for a schematic view):

\begin{figure*}
  \includegraphics[width=2\columnwidth]{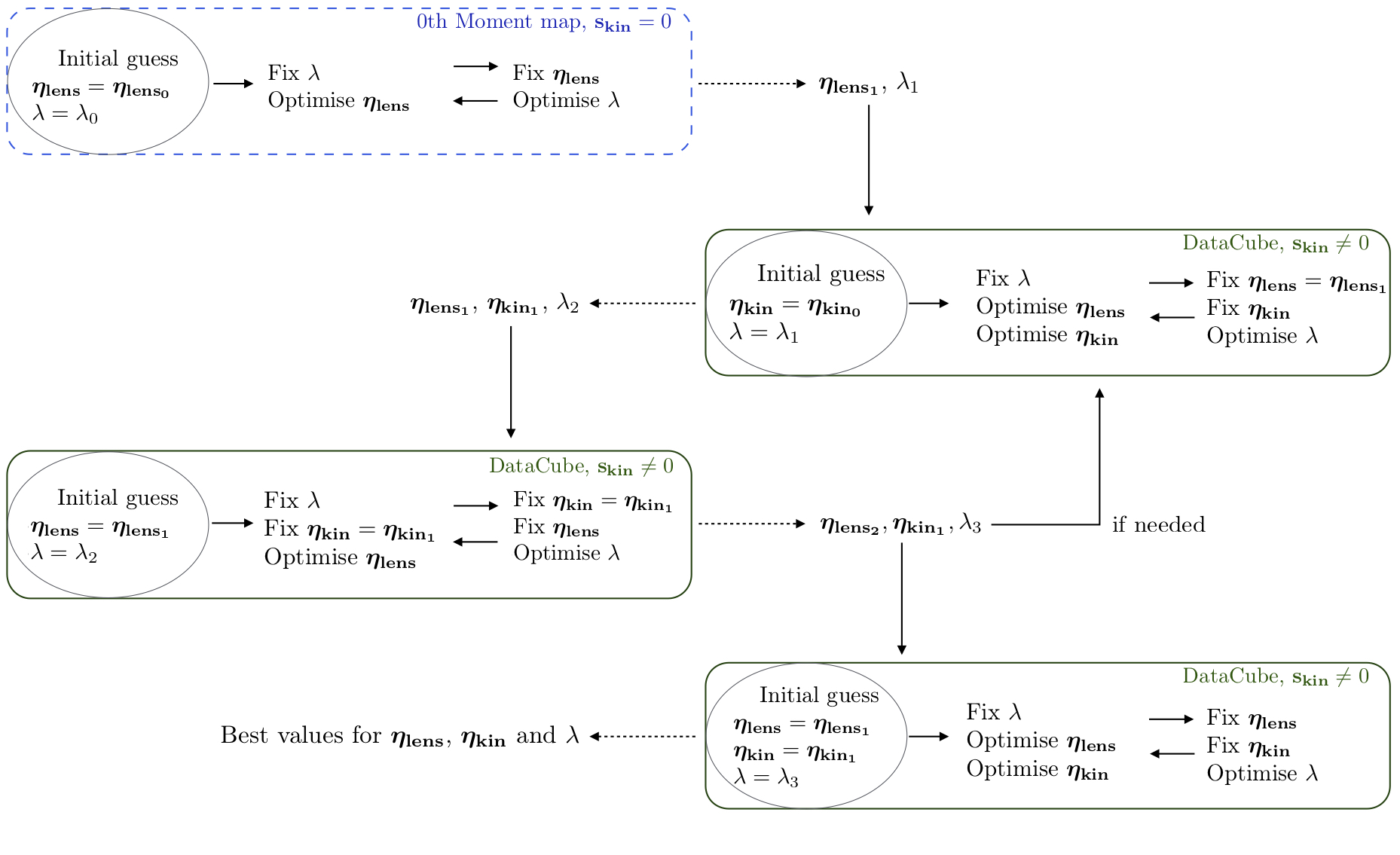}
  \caption{A schematic overview of the four-step optimisation scheme used to infer the unknown parameters $\boldsymbol{\eta_{\mathrm{lens}}}$, $\lambda$, $\boldsymbol{\eta_{\mathrm{kin}}}$. The four boxes represent the points \ref{item:1}-\ref{item:4} in Section \ref{sec:optimisation}. An initial estimate of the lens parameters is obtained by fitting the zero-th moment map, while for the successive steps the full 3D data cube is used. }
  \label{fig:flowchart}
\end{figure*}

\begin{enumerate}
\item To find a good initial guess for the lens model parameters, $\boldsymbol{\eta_{\mathrm{lens}}}$, we start by modelling the zeroth-moment map of the data. This optimisation is performed in three separate sub-steps. First, $\lambda$ is kept fixed at a relatively large value, such that the source model remains relatively smooth, and $P\left(\boldsymbol{\eta_{\mathrm{lens}}}|\lambda, \mathbf{d},\mathbfss{R}\right)$ is maximized relatively to $\boldsymbol{\eta_{\mathrm{lens}}}$. Second, the lens parameters are kept fixed at the most probable values found at the previous step, while $P\left(\lambda|\boldsymbol{\eta_{\mathrm{lens}}}, \mathbf{d},\mathbfss{R}\right)$ is optimized for the source regularization level $\lambda$. Finally, $P\left(\boldsymbol{\eta_{\mathrm{lens}}}|\lambda, \mathbf{d},\mathbfss{R}\right)$ is maximized again for the lens parameters with a source regularization level fixed to the most probable value determined in the previous stage. At every point of the non-linear mass-model optimization, the corresponding most probable source surface brightness distribution $\mathbf{s_{\rm MP}}$ is obtained by solving the linear system (\ref{eq:lin_k}) with $\mathbf{s_{kin}}=0$. 
\label{item:1}

\item We now model the entire 3D data cube. Assuming the values of $\boldsymbol{\eta_{\mathrm{lens}}}$ found in step \ref{item:1}, we infer the optimal regularization constant $\lambda$ and $\boldsymbol{\eta_{\mathrm{kin}}}$, maximising equation~(\ref{eq:p_lam}) by varying first the kinematic parameters that define $\mathbf{s_{kin}}$, then the source regularization level $\lambda$ and finally the kinematic parameters again. At this stage, the user can choose between a value of the regularization level $\lambda$ which is the same for all of the spectral channels or a value that varies channel by channel.
\label{item:2}

\item We repeat the process described in \ref{item:1}, using equations (\ref{eq:posterior_nl}) and (\ref{eq:lin_k}) with $\boldsymbol{\eta_{\mathrm{kin}}}$ equal to the value found in \ref{item:2}.
\label{item:3}

\item Finally, the lens parameters, $\lambda$ and only the kinematic parameters that describe the rotation velocity and velocity dispersion, i.e. $V_{\mathrm{rot}}$, $\sigma_{\mathrm{gas}}$, are simultaneously left free to vary, starting from the values of the parameters found at the previous steps. As for the last two, at this stage, we focus on the 3D data cube.\label{item:4}
\end{enumerate}

The analysis described at the points \ref{item:2} and \ref{item:3} are repeated if a visual inspection of the residuals reveals a mismatch between the model and the data.
All of the optimisation steps described above are done with a non-linear optimiser \citep[i.e. a Downhill-Simplex with Simulated Annealing,][]{press}. As discussed in Section \ref{sec:third},
the calculation of the Bayesian evidence with {\sc MultiNest} allows us to explore the parameter space and obtain the posterior distributions of the parameters. In this case, both the kinematic and lens parameters are simultaneously changed.
\section{IFU mock data}
\label{sec:mock}

To investigate the ability of our new modelling technique to recover reliable lens and kinematic parameters we simulate nine observations of H$\alpha$ emission from star-forming lensed galaxies at redshifts between $\sim$1.3 and $\sim$2.4. In particular, we use the technical features of the OSIRIS spectrograph \citep{larkin}. We have chosen to focus on OSIRIS because it has the typical characteristics of a near-infrared integral field spectrometer mounted on an 8-10m telescope in terms of spatial resolution, AO performances and spectral resolution, with a typical channel width of 30-40 km s$^{-1}$.

To simulate the lensed data we first build a cube from a rotating galaxy (Section \ref{sec:source_mod}), we then lens it forward using the lens mass model described in Section \ref{sec:lens}. Finally, we convolve the lensed cube with a spatial PSF and add the noise (Section \ref{sec:simulated}).\\

\subsection{Simulated sources}
\label{sec:source_mod}

The lensed sources have redshifts between 1.3 and 2.4 (column 2 in Table \ref{tab:filter}) which results in their H$\alpha$ emission line falling in the H or K filters (column 4 in Table \ref{tab:filter}). The total H${\alpha}$ fluxes (column 7 in Table \ref{tab:filter}) have values typical of star-forming galaxies at z$\sim$ 1 - 2 \citep[e.g.][]{forster2009, livermore}. The average resolving power is $\sim$3400 corresponding to $\sim$6$\AA$ in these bands. The cube of the rotating galaxy is built using $^{\mathrm{3D}}$BAROLO \citep{barolo}. Input values for the geometrical and kinematical parameters that define the inclination $i$ and position angle $PA$ and the rotation velocity $ V_{\mathrm{rot}}$ and dispersion $\sigma_{\mathrm{gas}}$ are listed in Table \ref{tab:filter}. The sources have an extension of $\sim$ 5 - 8 kpc along the major-axis, as typical of z$\sim$ 1 - 2 galaxies \citep[e.g.][]{wisnioski, leeth, genzel17, patricio}.

In the sections below, we provide more details on each model. In general, to check whether the functional forms in equations (\ref{eq:acrtg})-(\ref{eq:exp}) are flexible enough to reproduce a variety of realistic kinematical scenarios, we have considered different input rotation curves of varying complexity and different shapes. In particular, the mock data M1 and M2 are created and modelled with the same functional forms implemented in our code (Sections \ref{sec:m1}-\ref{sec:m2}). The mock data M3 is created with a different functional form (Section \ref{sec:m3}), while the simulated data M4, M5 and M6 have rotation curves derived from real observed galaxies (Sections \ref{sec:m4}-\ref{sec:m6}). The rotation curves of M1 and M4 are typical of dwarf galaxies, the rotation curves of M2 and M5 are prototypes of spirals, while those of M3 and M6 are typical of massive spirals with a prominent bulge. We have included dwarf galaxy kinematics to test if our code is able to recover the shape of the rotation curve when the turning point is not reached and only the increasing part is observable. Finally, the mock data M7, M8 and M9 are used to test the limits of our modelling technique. The aim is to quantify the minimum and maximum inclination angles that allow us to reliably recover the kinematics (M7 and M8, Sections \ref{sec:m7}-\ref{sec:m8}), as well as the minimum warp in the position angle that can be detected given the angular resolution of the data (M9, Sections \ref{sec:m9}).

\subsection{Simulated observations}
\label{sec:simulated}

We generate the simulated lensed data using a set of lens parameters $\boldsymbol{\eta_{\mathrm{lens}}}$ (see Table \ref{tab:filter}) that we have derived from the lens modelling of a set of real galaxy-scale lenses from the SLACS \citep{bolton} and SHARP \citep{lagatt} surveys. This choice is motivated by the fact that in this paper we are only focusing on galaxy-scale lenses. For the analysis of galaxy-cluster lenses more complicated mass distributions would have to be considered. We plan to investigate this issue further in a following paper. Using the lens equation, we lens forward the source surface brightness to the image plane for each spectral channel of the source cube. The simulated datasets have a spatial pixel scale in the image plane of 0.1 arcsec. Taking the OSIRIS characteristics, the field of view (FOV, column 5 in Table \ref{tab:filter}) varies between 3.6$\times$6.4 arcsec$^2$ and  4.8$\times$6.4 arcsec$^2$ depending on the filter (column 4 in Table \ref{tab:filter}). We convolve the lensed cube with a spatial PSF, $g$, which is described by the combination of two Gaussians 
\begin{equation}
 g=S\,g_{\mathrm{dif}}+\left(1-S\right)\,g_{\mathrm{seeing}}.
\end{equation}
This assumption allows us to take into account for the effects of the AO system, in the sense that the light of the source is divided between a diffraction limited-core, $g_{\mathrm{dif}}$, and a seeing-limited halo, $g_{\mathrm{seeing}}$, for a given strehl of the AO correction $S$  \citep{law06}. In particular, M1 to M3 and M8 are simulated using a  Full Width at Half Maximum (FWHM) of about 0.17 arcsec for $g_{\mathrm{dif}}$, a FWHM of about 0.95 arcsec for $g_{\mathrm{seeing}}$, and $S=0.2$; for M4 to M7 and M9 we use a FWHM equal to 0.2 arcsec and 0.6 arcsec for $g_{\mathrm{dif}}$ and $g_{\mathrm{seeing}}$, respectively, and $S=0.24$. All values of the PSF parameters are typical of OSIRIS observations \citep[e.g.][]{stark08, jones10, wisnioski11}. The effect of the spectral resolution is included on the plane of the source as described in Section \ref{sec:kinematic_model}.

To simulate a realistic noise distribution we create the sky-subtracted data, $\mathbf{d}$, using the simulation method by \citet{law06}, which was specifically designed for OSIRIS observations. The value of $\mathbf{d}$ is then obtained as
\begin{equation}
\mathbf{d}=\mathbf{n}+\mathbf{t}\,,
\end{equation}
where the noise $\mathbf{n}$ is a value extracted from a Gaussian distribution with a dispersion given by the sum in quadrature of the counts from the observed target, $\mathbf{t}$, and the background $\mathbf{t_{\mathrm{BG}}}$. For our mock observations $\mathbf{t}= \sum_{\mathrm{i}}^{\mathrm{Nch}} \mathbfss{M}\mathbfit{s}_{\mathrm{i}}$.

As explained in details by \citet{law06} the background count rate $\mathbf{t_{\mathrm{BG}}}$ is a function of the wavelength and takes into account the Mauna Kea near-IR sky brightness spectrum, the telescope emissivity and the AO system emissivity. We have taken into account the updated characteristics of the telescope and OSIRIS spectrograph relatively to those used by \citet{law06}: improved grating efficiency \citep[$\sim$0.78,][]{mieda14} and the halved read-out noise given by the installation of a new detector (T. Jones, private communications). The exposure times used for the simulated data-sets M1-M9 are listed in the eighth column of Table \ref{tab:filter} and they are typical of data containing star-forming lensed galaxies \citep{livermore, leeth}. The resulting mock data have a median SNR of $\sim$14 (see Figure \ref{fig:snrm1m9} in Appendix \ref{app:snr}).

 \begin{table*}
  \centering
  \caption{Observational and physical properties for the nine mock systems. \textbf{Top table}: Column 1: name of the dataset. Column 2: redshift of the source. Column 3: redshift of the lens. Column 4-5: OSIRIS filter and the corresponding FOV. Column 6: FWHM for the core+halo PSF (see Section \ref{sec:source_mod}). \textbf{Middle table}: Kinematic parameters used to create the source. \textbf{Bottom table}: Lens parameters used to lens the source and to create the observed mock data.}
  \label{tab:filter}
  \begin{tabular}{l l l l l l l l l l}
    \hline
    \multicolumn{10}{c}{Observation set-up}\\
    Mock dataset & $z_{\mathrm{source}}$ & $z_{\mathrm{lens}}$ & Filter & FOV & FWHM  & F(H$\alpha$) & t$_{\mathrm{exp}}$\\
           &  &  &  & \multicolumn{1}{c}{arcsec} & \multicolumn{1}{c}{arcsec} & 10$^{-18}$ erg s$^{-1}$ cm$^{-2}$ & ks\\
  \hline
    M1 & 2.05 & 0.881 & Kn1 & 3.6$\times$6.4 & 0.17+0.95 & 15 & 14.4\\
    M2 & 2.19 & 0.191 & Kn2 & 4.5$\times$6.4 & 0.17+0.95 & 20 & 14.4\\ 
    M3 & 2.15 & 0.722 & Kn2 & 4.5$\times$6.4 & 0.17+0.95 & 33 & 14.4 \\
    M4 & 2.26 & 0.191 & Kn3 & 4.8$\times$6.4 & 0.20+0.60 & 15 & 12.6\\
    M5 & 1.34 & 0.410 & Hn2 & 4.5$\times$6.4 & 0.20+0.60 & 6 & 10.8\\
    M6 & 2.36 & 0.881 & Kn3 & 4.8$\times$6.4 & 0.20+0.60 & 6 & 12.6\\
    M7 & 1.34 & 0.410 & Hn2 & 4.5$\times$6.4 & 0.20+0.60 & 9 & 10.8\\
    M8 & 2.19 & 0.191 & Kn2 & 4.5$\times$6.4 & 0.17+0.95 & 20 & 14.4\\
    M9 & 1.34 & 0.410 & Hn2 & 4.5$\times$6.4 & 0.20+0.60 & 10 & 12.4\\
    \\
    \hline 
   \multicolumn{10}{c}{Input kinematic parameters}\\
    Mock dataset & $i$ & $PA$ & $V_{\mathrm{t}}$ & $R_{\mathrm{t}}$ & $\beta$ & $\xi$ & $\sigma_{0}$ & $R_{0}$ & $\sigma_{1}$\\
    & $^\circ$ & $^\circ$ & km s$^{-1}$ & kpc &  &  & km s$^{-1}$ & kpc & km s$^{-1}$\\
  \hline
    M1 & 72.0 & 265.0 & 120.0& 2.0 & $\_$  &  $\_$  & 30.0 & -1.5 & $\_$\\
    M2 & 52.0 & 100.0 & 223.0 & 1.0  &  $\_$ &  $\_$   & 15.0 & 1.2 & 25.0\\
    M3 & 64.0 & 23.0  & 157.2 & 27.4 & 1.13 & 93.7 & 29.0 &  $\_$  &  $\_$ \\
    M4 & 59.0 & 145.0 & 73.7 & 5.52 & 0.24 & 50.1 & 46.0 & -1.19 & $\_$\\
    M5 & 68.0 & 280.0 & 151.4 & 2.17 & $\_$  & $\_$ & 34.0 & 26.0 & $\_$\\
    M6 & 65.0 & 45.0 & 219.7& 0.65 & 0.56 & 5.6 & 38.0 & $\_$ & $\_$\\
    M7 & 40.0 & 280.0 & 151.4 & 2.17 & $\_$  & $\_$ & 34.0 & 26.0 & $\_$\\
    M8 & 80.0 & 100.0 & 223.0 & 1.0  &  $\_$ &  $\_$   & 15.0 & 1.2 & 25.0\\
    M9 & 68.0 & 280.0/-3.75 & 151.4 & 2.17 & $\_$  & $\_$ & 34.0 & 26.0 & $\_$\\
    \\
    \hline
    \multicolumn{10}{c}{Input lens parameters}\\
    Mock dataset & $\kappa_0$ & $\theta$ & $q$ & $\gamma$ & $\Gamma_{\mathrm{sh}}$ & $\theta_{\mathrm{sh}}$ \\
      & arcsec & $^\circ$ &     &     &     &  $^\circ$\\
  \hline
    M1 & 1.44 & -12.72 & 0.82 & 2.06 & -0.039& 13.33  \\
    M2 & 1.33 & 157.95 & 0.93 & 2.28 & 0.050 & 174.45 \\
    M3 & 1.00 & 0.00   & 0.99 & 2.00 & 0.240 & 38.00   \\
    M4 & 1.33 & 157.95 & 0.93 & 2.28 & 0.050 & 174.45 \\
    M5 & 0.81 & 71.20  & 0.84 & 2.00 & 0.096 & 34.40  \\
    M6 & 1.44 & -12.72 & 0.82 & 2.06 & -0.039& 13.33  \\
    M7 & 0.81 & 71.20  & 0.84 & 2.00 & 0.096 & 34.40  \\
    M8 & 1.33 & 157.95 & 0.93 & 2.28 & 0.050 & 174.45 \\  
    M9 & 0.81 & 71.20  & 0.84 & 2.00 & 0.096 & 34.40  \\
    \\
    \hline
    \end{tabular}
\end{table*}

\section{Modelling strategy}
\label{sec:modelling}

In this section, we describe how we build the kinematic prior $\mathbf{s_{kin}}$ and derive the best kinematic parameters $\boldsymbol{\eta_{\mathrm{kin}}}$ (we refer to Section \ref{sec:kinematic_model} for a definition). In particular, we discuss the assumptions made to define the radial extension $R_{\mathrm{ext}}$, the centre and the systemic velocity and the initial conditions for the geometrical and kinematic parameters for the specific data analysed in this paper. These assumptions can change depending on e.g. the data quality of the observations, previous estimates of the kinematic and/or geometric parameters, the accuracy of the redshift of the source.

The first step is to define the radial extension and the effective resolution on which $\mathbf{s_{kin}}$  is sampled. From the reconstruction of the zeroth moment, we first derive a SNR map on the reconstructed source, by propagating the observational noise of the data. We then define the radial extension $R_{\mathrm{ext}}$ of the kinematic model as the radius along the apparent major axis of the galaxy at which the SNR$\sim$3. The kinematic models are built using a ring width that is half the size of the pixels on the image plane. We have explicitly verified that these choices do not influence the recovered kinematic parameters. The Cartesian grid is then mapped onto a triangular adaptive grid, with triangles of average dimensions between $\sim$10$^{-3}$ arcsec to $\sim$10$^{-1}$ arcsec (this is set by a combination of the pixel scale on the image plane and the lensing magnification).

To reduce the number of free kinematic parameters during the optimisation, we chose to keep the centre of the source galaxy fixed at the flux-weighted average position of the zeroth moment map (these differ by at most by 1 per cent from the correct values). The systemic velocity is also kept fixed at zero km s$^{-1}$. When dealing with real data, one will be able to estimate its value from the global velocity profile, where the latter is obtained from the source intensity in each spectral channel of the data cube or other independent estimations. In Section \ref{sec:further_tests} we discuss the results obtained by changing the centre and systemic velocity from the true values. 
The free kinematic parameters are then $\boldsymbol{\eta_{\mathrm{kin}}}=\{i, PA, V_{\mathrm{rot}}, \sigma_{\mathrm{gas}}\}$.

Since the geometrical and kinematic parameters are coupled and degenerate (see equation \ref{eq:vlos}), they need to be initialised with educated guess values. In this paper, we estimate the geometrical parameters ($i$ and $PA$) by applying $^{\mathrm{3D}}$BAROLO to the 3D source derived from the lens parameters inferred at point \ref{item:1} of Section \ref{sec:sourcerec}. We set the initial values for $V_{\mathrm{t}}$ and $R_{\mathrm{t}}$ that define the rotation curve to the arbitrary, but observationally motivated, values of 100 km/s and 1 kpc respectively \citep[e.g.][]{livermore, jones10}. For the multi-parameter function, we set $\beta=0.2$ and $\xi=10.0$ as initial guesses. The choice of the functional form is arbitrary, but it should be noted that the multi-parameter function is the most flexible one and it reproduces the arctangent function for $\xi=1.1$. Furthermore, as demonstrated in Section \ref{sec:m2}, a wrong choice of the functional form for the rotation curve leads to systematic image residuals, indicating that a different choice should be made. The initial value for $\sigma_0$ is set to 30 km/s, while initial guesses for the other parameters that define the velocity dispersion functions, are chosen such that $\sigma\left(R_{\mathrm{ext}}\right)-\sigma_0$ is not larger than 20 km/s, as it is typical for the ionized gas of star-forming galaxies \citep[e.g.][]{epinat10, diteodoro16, mason}.

\subsection{Functional forms for the rotation velocity}
Here, we briefly summarize the functional forms used to create and model the rotation velocities that define $\mathbf{s_{kin}}$ (see the second and third columns in Table \ref{tab:summary}). For the background galaxy of the simulated data M1, we assume a hyperbolic tangent function for the rotation velocity (blue squares in Figure \ref{fig:vel}). The data are then modelled assuming the same parametric form that we have used to create them. These simulated data represent, therefore, a zeroth-order test of our modelling technique.

The mock data M2 were built assuming an arctangent function for the rotation velocity (blue squares in Figure \ref{fig:vel}). The data are modelled twice, once with the same functional form that we have used as input and once with a hyperbolic tangent function.

We have created the simulated data M3 using the following functional form for the rotation velocity (blue squares in Figure \ref{fig:vel}):
\begin{equation}
V_{\mathrm{rot}}\left(R\right)=\sqrt{V_{1}^2+V_{2}^2},
\label{eq:rc_s3}
\end{equation}
where $V_1$ is the contribution from an isothermal dark-matter halo while $V_{2}$ represents the contribution from a S\'ersic profile with a S\'ersic index $n=1$ \citep{freeman}. 
Since our lens modelling code does not include the functional form expressed by equation (\ref{eq:rc_s3}), we model these mock data using the flexible multi-parameter function (dashed red line in Figure \ref{fig:vel}). After checking that the input rotation curve is well reproduced (see discussion in Section \ref{sec:m3} for further details), we fit the input 1D rotation curve with the multi-parameter function (blue solid line in Figure \ref{fig:vel})\footnote{The fitting of the input rotation curves for M3 to M6 was done using the Python package Scipy.optimize}. By comparing the results of this fit with the kinematic parameters derived by the 3D lens modelling code, we can then quantify the accuracy of our technique and study the systematic errors that may derive from the choice of the kinematic functional forms.

We have created the mock datasets M4, M5 and M6 using the rotation curves measured for three low-redshift galaxies: NGC\,2976, NGC\,3198 and NGC\,6674 from \citet{lelli}.
For these sources the values of the rotation parameters listed in Table \ref{tab:filter} are obtained by fitting the input data points (blue squares in Figure \ref{fig:vel}) with one of the functional forms implemented in our code (blue solid line in Figure \ref{fig:vel}). As for M2, this fitting allows us to evaluate an accuracy of the inferred kinematic parameters which is independent of the choice of the parametrisation. As discussed in Sections \ref{sec:m4} to \ref{sec:m6} these mock data are then modelled assuming the multi-parameter function for M4 and M6 and the hyperbolic tangent function for M5.

The simulated data M7, M8 and M9 are built to test the limits of our method. M7 and M8 have the same kinematics of M5 and M2 but an inclination angle of 40$^\circ$ and 80$^\circ$, respectively. M9 has the same kinematics of M5, but it has a strong warp, which causes a change of 30$^\circ$ of its position angle across the galaxy. 

 \begin{table*}
  \centering
  \caption{For each model in column 1 we show the assumptions on the input (second column) and recovered (third column) shapes for the rotation curves. The fourth and fifth columns show the median uncertainties on the lens and kinematic parameters. The sixth and seventh columns show the relative accuracies for the lens and kinematic parameters respectively.}
  \label{tab:summary}
  \begin{tabular}{c c c c c c c}
    \hline
    Mock dataset & Input RC & Model RC & Median uncertainty & Median uncertainty & Median accuracy  & Median accuracy \\
 & &  &  on $\boldsymbol{\eta_{\mathrm{lens}}}$ & on $\boldsymbol{\eta_{\mathrm{kin}}}$ &  on $\boldsymbol{\eta_{\mathrm{lens}}}$  & on $\boldsymbol{\eta_{\mathrm{kin}}}$\\   
           &  &  & $\%$ &$\%$ &$\%$ & $\%$\\
  \hline
    M1 & Arctangent & Arctangent & 2 & 7 & 3.6 & 2.3 \\
    M2 & Hyperbolic & Hyperbolic & 5 & 13 & 1.3 & 2.8\\
    M3 & Equation (\ref{eq:rc_s3}) & Multi-parameter &  5 & 3 & 2.8 & 1.7\\
    M4 & NGC\,2976 & Multi-parameter & 5 & 9 & 1.5 & 3.0 \\
    M5 & NGC\,3198 & Hyperbolic & 3 & 8 & 0.5 & 0.5 \\
    M6 & NGC\,6674 & Multi-parameter & 6 & 9 & 1.0 & 1.1\\
    M7 & NGC\,3198 & Hyperbolic & 8 & 6 & 2.4 & 3.1\\
     & +low inclination\\
    M8 & Hyperbolic & Hyperbolic & 3 & 7 & 0.7 & 2.0 \\
     & +large inclination\\
    M9 & NGC\,3198 & Hyperbolic & 6 & 7 &1.9 & 0.5 \\
     & +warp\\
    \\
    \hline 

    \end{tabular}
\end{table*}

\section{Results}
\label{sec:results}

To test the ability of our method to recover the input lens and kinematic parameters, we model the nine mock dataset introduced in Section \ref{sec:simulated} with the new modelling technique described in Section \ref{sec:sourcerec}. All assumptions made during the modelling procedure were discussed in Section \ref{sec:modelling}.

We obtain the uncertainties on the inferred parameters using {\sc MultiNest} (see Section \ref{sec:optimisation}). For each parameter we adopt priors which are flat in the intervals $[\boldsymbol{\overline{\eta}_{\mathrm{lens/kin}}}-0.2\boldsymbol{\overline{\eta}_{\mathrm{lens/kin}}}, \boldsymbol{\overline{\eta}_{\mathrm{lens/kin}}}+0.2\boldsymbol{\overline{\eta}_{\mathrm{lens/kin}}}]$, where $\boldsymbol{\overline{\eta}_{\mathrm{lens/kin}}}$ are the best-fit parameters, inferred from the non-linear optimization (see Section \ref{sec:sourcerec}). To be conservative, we report as errors in the parameters the sum in quadrature of the following two contributions: the 1-$\sigma$ uncertainties on the posterior distributions derived by {\sc MultiNest} and the difference between the maximum a posteriori parameter values obtained by {\sc MultiNest}, and the non-linear optimizer. This difference arises because, as discussed above, the geometrical parameters that describe the source, i.e. $PA$ and $i$, are kept fixed during the optimisation, while they are left free to vary during the {\sc MultiNest} exploration of the parameters space. In most cases, the discrepancy is smaller than 5 per cent, while it reaches a level of 20 per cent in one case that will be discussed separately (see Section \ref{sec:m6}). For the mock data M1, M2 and M8, for which we have used the same functional forms to create and model the data, these errors only account for the statistical uncertainties, while for the other models they also provide an estimate of the systematic errors, related to the choice of parametrisation. The median relative uncertainties on $\boldsymbol{\eta_{\mathrm{lens}}}$ and $\boldsymbol{\eta_{\mathrm{kin}}}$ for each model are listed in Table \ref{tab:summary} (fourth and fifth columns respectively), while the sixth and seventh columns in Table \ref{tab:summary} list the relative accuracies\footnote{The median accuracies on $\boldsymbol{\eta_{\mathrm{lens}}}$ and $\boldsymbol{\eta_{\mathrm{kin}}}$ are calculated taking into account the relative difference between the input (Table \ref{tab:filter}) and recovered parameters (Table \ref{tab:output}) as: (input-recovered)/input.}.

\begin{figure*}
        \includegraphics[width=1.5\columnwidth]{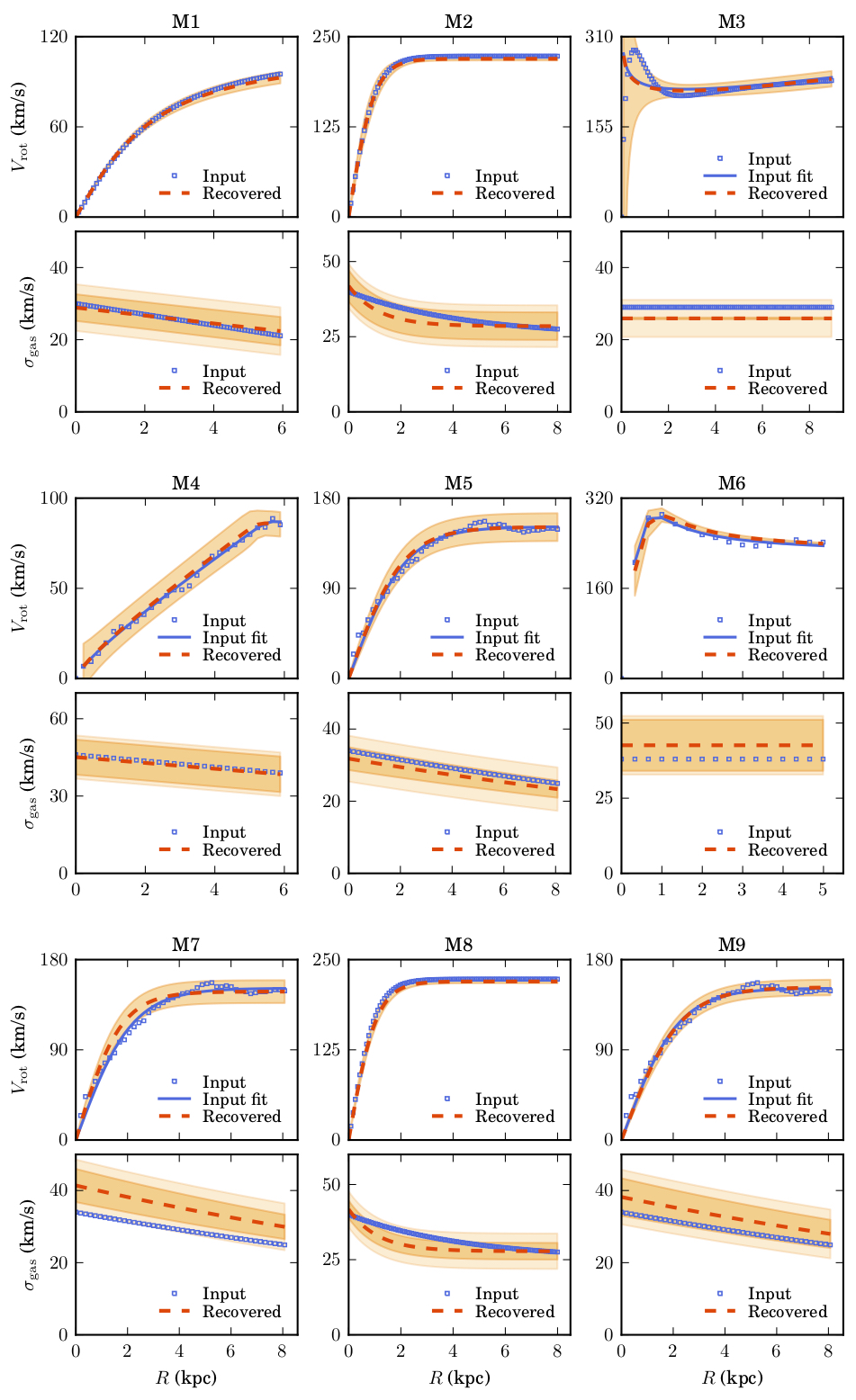}
        \caption{Rotation curves and velocity dispersions for the mock dataset M1-M9. The blue squares are the input rotation curves and velocity dispersion profiles used to create the cubes containing the sources. These are then lensed forward to build the lensed  mock data. The dashed red lines are the functional forms that best describe the kinematic priors, while the solid blue line for M3-M7 and M9 show the fit to the input data using the same functional forms as those used for the kinematic priors at the 3D lens modelling stage. The orange bands for $V_{\mathrm{rot}}$ and $\sigma_{\mathrm{gas}}$ are obtained by error propagation from the uncertainties of the parameters that defined the rotation curves and velocity dispersion profiles, while the light orange bands for $\sigma_{\mathrm{gas}}$ take into account also the contribution from the spectral resolution (see Section \ref{sec:results} for further details). In the velocity dispersion profile of M3, the orange band is too thin (0.25 km  s$^{-1}$) to be visible, see discussion in Section \ref{sec:m3} for further details.}
        \label{fig:vel}
\end{figure*}

\begin{figure*}
        \includegraphics[width=2\columnwidth]{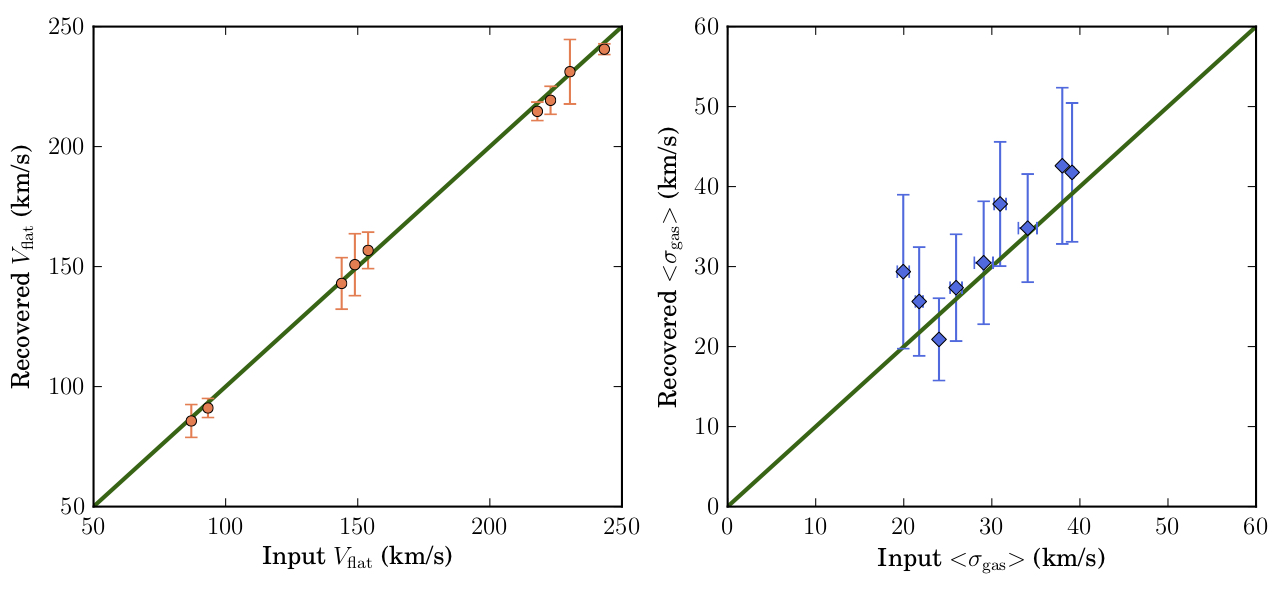}
        \caption{\textit{Left:} Recovered versus input values of $V_{\mathrm{flat}}$ for the nine mock datasets. Some points are shifted both on x and y axis by the same quantity for a better visualisation of all the points. The green line represents a 1:1 relation between quantities on x and y axis. \textit{Right:} same as in the left panel, but for $\mean{\sigma_{\mathrm{gas}}}$. The errors bars take into account both the standard uncertainties due to error propagation and the standard deviation due to fact that these are averaged quantities.}
        \label{fig:vflat}
\end{figure*}

Figures \ref{fig:m1} (for M1) and \ref{fig:m2}-\ref{fig:m9} in Appendix \ref{sec:appa} (for M2-M8) show the comparison between mock observations and best-fit models. We plot the contour levels of the input source (first column), the simulated lensed data (second column), the inferred lensed model (third column), the normalised image residuals (fourth column), the reconstructed source (fifth column) and the contour levels of the kinematic model (sixth column), for a selected number of spectral channels. We present the input and recovered rotation curves and velocity dispersion profiles with blue squares and dashed red lines respectively in Figure \ref{fig:vel}. The orange band for the rotation velocities denotes the uncertainties $\epsilon_{\mathrm{p}}$, obtained after propagating the uncertainties on the parameters that describe the profiles (see Table \ref{tab:output}). To take into account the contribution to the velocity dispersion uncertainties due to the spectral resolution, we compute the uncertainties on the values of $\sigma(R)$ as $\sqrt{\epsilon_{\mathrm{p}}^2+\epsilon_{\mathrm{cw}}^2}$ (light orange bands in Figure \ref{fig:vel}). $\epsilon_{\mathrm{p}}$ (orange bands in Figure \ref{fig:vel}) has the same meaning defined above for $V_{\mathrm{rot}}(R)$, while $\epsilon_{\mathrm{cw}}$ is obtained as the FWHM of the channel width divided by 3$\times$2.355. The factor of 3 is obtained after testing the effect of the spectral resolution on the recovered velocity dispersion with mock data. We list the inferred lens and kinematic parameters in Table \ref{tab:output}. These values should be compared with those reported in Table \ref{tab:filter}. \\
In Figure \ref{fig:vflat} we show the comparison between the input and recovered flat velocities $V_{\mathrm{flat}}$ and average velocity dispersions $\mean{\sigma_{\mathrm{gas}}}$. The values of $V_{\mathrm{flat}}$ are obtained as the average of the rotation velocities at large radii, while the $\mean{\sigma_{\mathrm{gas}}}$ are obtained by averaging the values of $\sigma_{\mathrm{gas}}(R)$ starting from $R=0$ kpc. The error bars in Figure \ref{fig:vflat} take into account the contribution of both the uncertainties on the values of $V_{\mathrm{rot}}$ and $\sigma_{\mathrm{gas}}$ at each radius, as showed by the orange and light orange bands in Figure \ref{fig:vel}, and the standard deviation. The flat part of the rotation curves is correctly reproduced for all mock dataset. In particular, with our technique we are able to recover $V_{\mathrm{flat}}$ not only for the galaxies for which the input rotation curves is described by functional forms, but also for the rotation curves taken from real galaxies. This test ensures us that the functional forms implemented in our code are flexible enough to reproduce the variety of shape of rotation curves (from dwarf to massive galaxies). Even if the details in the inner region could be lost (see Section \ref{sec:m3}), the physical parameters that depend on $V_{\mathrm{flat}}$ can be exactly recovered. The values of $\mean{\sigma_{\mathrm{gas}}}$ are recovered with a great accuracy, even if they are more affected by the spectral resolution.

Below, we provide a detailed discussion on the modelling results for each of the nine simulated datasets. The reader not interested in the details can skip to Section \ref{sec:further_tests}, where we address some key issues related to radial motions, signal-to-noise ratios and changes in the centre coordinates and systemic velocities and to Section \ref{sec:conclusions} where we summarise the results of our tests.

\begin{table*}
  \centering
  \caption{\textbf{Top table}: Recovered kinematic parameters that best describe the prior for the nine sources. \textbf{Bottom table}: Recovered lens parameters for the nine mock datasets. The uncertainties are derived using a {\sc MultiNest} method, as explained in Section \ref{sec:results}.}
  \label{tab:output}
  \begin{tabular}{l l l l l l l l l l}
    \hline
     \multicolumn{10}{c}{Recovered kinematic parameters}\\
    Mock dataset & $i$ & $PA$ & $V_{\mathrm{t}}$ & $R_{\mathrm{t}}$ & $\beta$ & $\xi$ & $\sigma_{0}$ & $R_{0}$ & $\sigma_{1}$\\
    & $^\circ$ & $^\circ$ & km s$^{-1}$ & kpc &  &  & km s$^{-1}$ & kpc & km s$^{-1}$\\
  \hline
    M1 & 74.0$\pm$1.6 & 260.0$\pm$3.2 & 115.0$\pm$4.8 & 1.86$\pm$0.23  &  $\_$  &  $\_$ & 28.9$\pm$3.7 & -1.1$\pm$0.23 & $\_$\\
    M2 & 54.4$\pm$0.1 & 102.6$\pm$2.7 & 219.3$\pm$2.2 & 0.97$\pm$0.09 & $\_$ & $\_$ & 13.4$\pm$2.6 & 1.15$\pm$0.21 & 28.5$\pm$4.6 \\
    M3 & 63.2$\pm$0.4 & 24.6$\pm$1.5  & 155.4$\pm$4.2 & 27.1$\pm$5.2 & 1.09$\pm$0.02 & 95.8$\pm$13.1 & 25.9$\pm$0.25 & $\_$  &  $\_$\\   
    M4 & 60.0$\pm$1.8 & 150.0$\pm$5.6 & 72.9$\pm$5.6 & 5.23$\pm$0.53 & 0.25$\pm$0.03 & 51.4$\pm$5.3 & 43.9$\pm$6.8 & -1.13$\pm$0.20 & $\_$\\
    M5 & 70.7$\pm$5.9 & 282.9$\pm$2.5 & 151.2$\pm$13.9 & 2.07$\pm$0.18 &  $\_$  &  $\_$ &  31.8$\pm$3.2 & 26.2$\pm$2.1 & $\_$\\
    M6 & 62.0$\pm$3.3 & 45.0$\pm$4.2 & 220.9$\pm$1.2 & 0.75$\pm$0.13 & 0.57$\pm$0.03 & 4.80$\pm$1.70 & 42.6$\pm$8.5  & $\_$ & $\_$\\
    M7 & 40.2$\pm$0.5 & 281.4$\pm$2.7 & 151.2$\pm$11.4 & 2.09$\pm$0.13 &  $\_$  &  $\_$ &  41.1$\pm$4.6 & 24.9$\pm$1.8 & $\_$\\
    M8 & 81.2$\pm$2.5 & 97.5$\pm$3.8 & 219.7$\pm$2.4 & 1.07$\pm$0.09 & $\_$ & $\_$ & 13.9$\pm$1.0 & 1.12$\pm$0.12 & 27.8$\pm$2.8 \\
    M9 & 67.6$\pm$6.3 & 277.8$\pm$2.7/-3.3$\pm$1.9 & 152.4$\pm$7.8 & 2.23$\pm$0.18 &  $\_$  &  $\_$ &  38.2$\pm$5.8 & 25.9$\pm$2.6 & $\_$\\

    \\
    \hline
    \multicolumn{10}{c}{Recovered lens parameters}\\
    Mock dataset & $\kappa_0$ & $\theta$ & $q$ & $\gamma$ & $\Gamma_{\mathrm{sh}}$ & $\theta_{\mathrm{sh}}$ \\
      & arcsec & $^\circ$ &     &     &     &  $^\circ$\\
    \hline
    M1 & 1.43$\pm$0.01 & -15.6$\pm$0.5 & 0.79$\pm$0.02 & 2.05$\pm$0.01 & -0.044$\pm$0.004 & 10.2$\pm$0.8\\
    M2 & 1.34$\pm$0.05 & 155.8$\pm$4.9 & 0.95$\pm$0.06 & 2.25$\pm$0.10 & 0.056$\pm$0.011 & 173.38$\pm$0.11\\
    M3 & 0.97$\pm$0.04 & -0.08$\pm$0.01  & 0.98$\pm$0.01 & 2.04$\pm$0.06 & 0.257$\pm$0.025  & 39.1$\pm$0.07 \\
    M4 & 1.31$\pm$0.09 & 160.50$\pm$6.5 & 0.95$\pm$0.01 & 2.31$\pm$0.16 & 0.051$\pm$0.009 & 173.8$\pm$6.8\\
    M5 & 0.78$\pm$0.03 & 69.6$\pm$0.4 & 0.83$\pm$0.06 & 2.03$\pm$0.04 & 0.096$\pm$0.002 & 34.3$\pm$3.2  \\
    M6 & 1.43$\pm$0.01 & -15.65$\pm$2.7 & 0.82$\pm$0.03 & 2.08$\pm$0.12 & -0.037$\pm$0.001 & 14.76$\pm$1.19 \\
    M7 & 0.83$\pm$0.04 & 72.1$\pm$6.2 & 0.81$\pm$0.07 & 1.96$\pm$0.09 & 0.093$\pm$ 0.002 & 33.3$\pm$3.1 \\
    M8 & 1.32$\pm$0.02 & 153.7$\pm$4.5 & 0.96$\pm$0.03 & 2.28$\pm$0.02 & 0.056$\pm$0.004 & 173.2$\pm$4.1\\
    M9 & 0.82$\pm$0.07 & 72.6$\pm$4.4 & 0.82$\pm$0.02 & 1.99$\pm$0.13 & 0.101$\pm$0.004 & 33.9$\pm$1.7\\

    \\
    \hline
    \end{tabular}
\end{table*}

\subsection{Mock dataset M1} 
\label{sec:m1}

The simulated data M1 were created assuming an arctangent function for the rotation velocity and a dispersion curve which is linearly declining from a value of 30\,km s$^{-1}$ at $R=0$\,kpc to 21\,km s$^{-1}$ at $R=5.9$\,kpc. The source position angle also changes linearly from 270$^\circ$ in the inner regions to 260$^\circ$ in the outer areas.\\
We model the simulated data with the same functional forms used to create them. Since the small change in the position angle is not detectable by a visual analysis of the zeroth-moment map, resulting from the first step of the optimisation scheme (see Section \ref{sec:optimisation}), we decided to keep it fixed to the constant value of 260$^\circ$ during the following steps. We have found that this assumption does not significantly influence the derived kinematics. 
The inferred parameters that define the rotation velocity and the velocity dispersion have median relative uncertainties of 7 per cent and are within 2-$\sigma$ from the input values. The inferred lens parameters, characterised by median relative uncertainties of 2 per cent, are within 1-$\sigma$ from the input values, with the only exceptions of the lens and external-shear position angles $\theta$ and $\theta_{\mathrm{sh}}$ which differ by 5.7-$\sigma$ and 3.9-$\sigma$, respectively, from the input values. This result is related to the fact that the lens is very close to being spherical and the shear strength almost negligible. 

\begin{figure*}
\includegraphics[width=2\columnwidth]{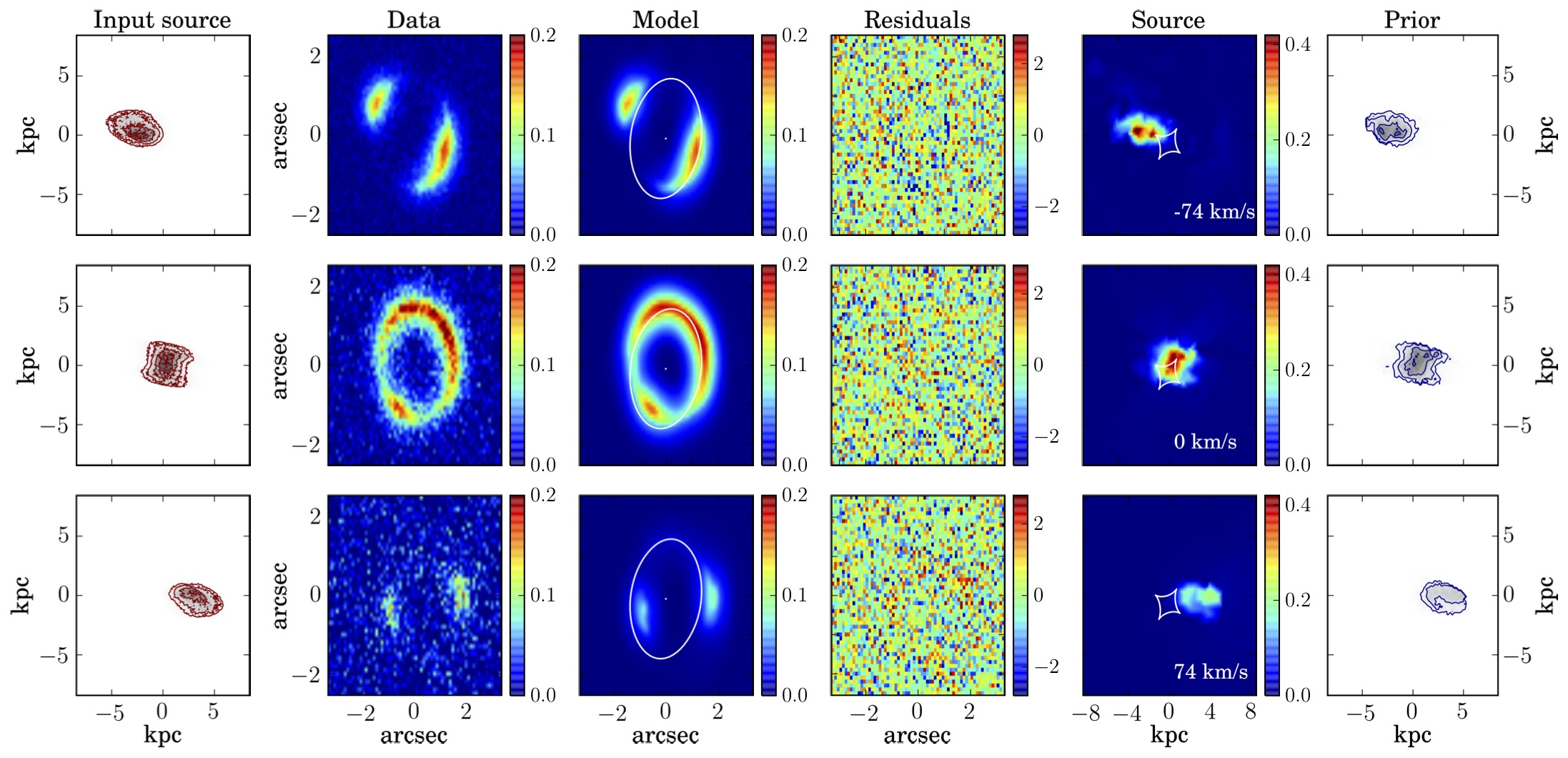}
\caption{The rows show some representative channel maps for the simulated dataset M1. Column 1: Input source. Column 2: Mock lensed data. Column 3: Lensed model and the corresponding critical curves. Column 4: Normalized residuals obtained as the ratio between the difference of the data and the model and the corresponding noise. Column 5: Reconstructed source and caustics. Column 6: Kinematic prior used to constrain the source reconstruction. The contour levels in the first and sixth columns are set at $n=0.1, 0.2, 0.4, 0.6, 0.8$ times the value of the maximum flux.}
  \label{fig:m1}
\end{figure*} 

\subsection{Mock dataset M2} 
\label{sec:m2}

\begin{figure}
\includegraphics[width=\columnwidth]{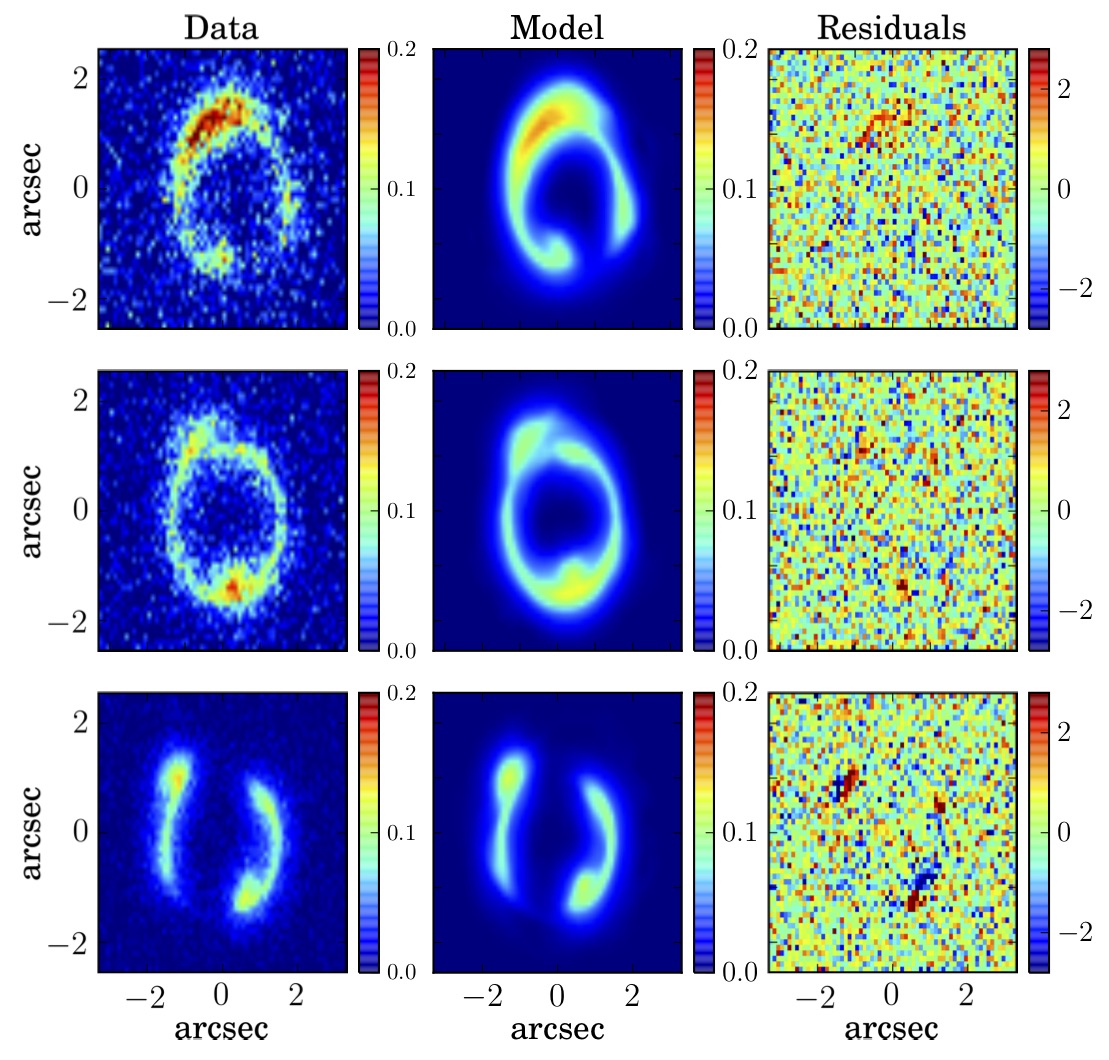}
\caption{Same as Figure \ref{fig:m2}, for a rotation velocity described by an arctangent function.}
\label{fig:m1arc}
\end{figure}

We have created the simulated data M2 using a hyperbolic tangent function for the source rotation curve and an exponential function for its velocity dispersion.\\
First, we model this dataset assuming the same functional forms used as input. This choice produces normalised residuals that are of the order of 0.5 per cent (see the fourth column in Figure \ref{fig:m2}). The inferred lens parameters have a median relative uncertainty of 5 per cent, while the recovered kinematic parameters have median relative uncertainties of 13 per cent (Table \ref{tab:output}). The largest contribution to the kinematic uncertainties comes from the velocity dispersion parameters, due to the limited spectral resolution (channel width of $\sim$36.8 km s$^{-1}$) of these data (see the orange band in Figure \ref{fig:vel}). The input lens and the kinematic parameters are within the 1-$\sigma$ uncertainties from the recovered values. 
To test our capability to distinguish between different forms of parametrisation, we have also modelled this dataset with an arctangent function. We have found that under this assumption the residuals get worse (see the third column in Figure \ref{fig:m1arc}), reaching a maximum value of $ \sim$6-$\sigma$. We have then computed the marginalised Bayesian evidence to compare and rank these two models. As anticipated in Section \ref{sec:optimisation}, the marginalized evidence in equation (\ref{eq:integ}), allows us to quantify how well a model m$_\mathrm{i}$ is supported by the data against another model m$_\mathrm{j}$. This quantification is expressed in terms of the Bayes factor, $\Delta\log E_{\mathrm{ij}}=\log E_\mathrm{i}-\log E_\mathrm{j}$. We find that the Bayes factor is of the order of 1400 against the arctangent model, meaning that the hyperbolic tangent function for the rotation curve is largely supported by the data. We can conclude, therefore, that the data contain enough information for us to infer the most suitable shape. 

\subsection{Mock dataset M3} 
\label{sec:m3}

The lens system M3 was created assuming a rotation curve for the background galaxy described by the functional form in equation (\ref{eq:rc_s3}). This function is not implemented in our code. The velocity dispersion was assumed to be constant.\\
We model these simulated data using the multi-parameter function given by equation (\ref{eq:mult}), which is the most flexible function that we have implemented. We find that the lens parameters are recovered with a median relative uncertainty of 5 per cent and are within 1-$\sigma$ from the input values. Our constraints on the lens mass model are therefore not significantly influenced by our assumptions on the source prior. The inferred parameters that define the rotation curve ($V_{\mathrm{t}}$, $R_{\mathrm{t}}$, $\beta$, $\xi$) in Table \ref{tab:output} should be compared with those reported in Table \ref{tab:filter}, obtained by fitting the input 1D rotation curve (blue squares in Figure \ref{fig:vel}) using the multi-parameter function (solid blue line in Figure \ref{fig:vel}). The inferred kinematic parameters have median uncertainties of 3 per cent, and they are within 2-$\sigma$ from the fitted values. The only exception is the recovered velocity dispersion which is more than 3-$\sigma$ away. However, this discrepancy is due to an underestimation of the uncertainties that do not include the systematic errors introduced by the spectral resolution. If we take into account the uncertainties due to spectral resolution, $\epsilon_{\mathrm{cw}}$ in Section \ref{sec:results}, the recovered dispersion profile is in agreement within 1-$\sigma$ with the input profile (see the light orange band in Figure \ref{fig:vel}).
We find that both the fit to the input rotation curve (solid blue line in Figure \ref{fig:vel}) and the rotation curve derived from our lens modelling technique (red dashed line in Figure \ref{fig:vel}) are a poor description of the inner regions of the real curve (see the blue squares in Figure \ref{fig:vel}).
Despite its flexibility, the multi-parameter function does not allow us to correctly reproduce the peculiarity of the data in the central regions. Correspondingly, we find that the overall fit to the data has systematic residuals that reach maximum values of $\sim$5-$\sigma$ (see the third column in Figure \ref{fig:m3}). However, the rotation velocity at the outer regions is well reproduced, ensuring that even if the details of the inner regions are lost, the physical parameters that depend on the asymptotic velocity (e.g. angular momentum and dynamical mass) can still be recovered with good accuracy (see Figure \ref{fig:vflat}). \\

\subsection{Mock dataset M4} \label{sec:m4}

The input values of the rotation velocity for this system are taken from the rotation curve of the nearby dwarf galaxy NGC\,2976 \citep{lelli}. This choice allows us to test whether the assumed functional forms are good enough to reproduce real rotational curves. A linear equation describes the velocity dispersion curve.\\
During the modelling phase, we use the multi-parameter function, equation (\ref{eq:mult}), to describe the rotation velocity, while for the velocity dispersion we use the same functional form used as input.  As for the simulated data M3, to have a quantitative estimate of the accuracy on the derived kinematics, we first fit the input 1D rotation curve with the same functional form used in the 3D lens modelling process (solid blue line in Figure \ref{fig:vel}). The recovered kinematic parameters have a median relative uncertainty of 9 per cent, while the lens parameters have a median relative uncertainty of 5 per cent (Table \ref{tab:output}). The inferred lens parameters are within the 2-$\sigma$ errors from the input values. The kinematic parameters are within 1-$\sigma$ from the values derived by fitting the 1D rotation curve.

\subsection{Mock dataset M5} \label{sec:m5}

As for the simulated dataset M4, we create M5 using the rotation curve of a real galaxy as input, in this case NGC\,3198 \citep{lelli}. The input functional form for the velocity dispersion is an exponential function, equation (\ref{eq:exp}), with $\sigma_1=0.0$ km s$^{-1}$. \\
At the modelling stage, we use the hyperbolic tangent function for the rotation curve and an exponential function with  $\sigma_1$ fixed at zero km s$^{-1}$ for the velocity dispersion. As for the simulated data M3 and M4, we start by fitting the 1D rotation curve with the same functional form used for the 3D lens modelling. 
We find that the hyperbolic tangent function provides a good enough description of the data. The normalised residuals (fourth column in Figure \ref{fig:m5}), indeed, do no show any systematic features, usually indicative of wrong assumptions in the building of the prior (e.g. see the model M2 and M3 in Sections \ref{sec:m2}, \ref{sec:m3}). The recovered lens and kinematic parameters have median relative uncertainties of 3 and 8 per cent respectively and they are within 1-$\sigma$ from the input values.
 
\subsection{Mock dataset M6} \label{sec:m6}

The simulated data M6 were created using the rotation curve of the nearby galaxy NGC\,6674 \citep{lelli}, while for the velocity dispersion curve we have used a constant value of 38 km s$^{-1}$.  When modelling the data, the prior is built assuming the multi-parameter function for the rotation curve, while the dispersion is assumed to be constant.\\
The input lens parameters (Table \ref{tab:filter}) are within the 1-$\sigma$ uncertainties from the recovered values (Table \ref{tab:output}). The kinematic parameters inferred by the 3D lens modelling technique are within 1-$\sigma$ from the values derived by fitting the 1D rotation curve directly.
We find that the inferred lens and kinematic parameters have a median relative uncertainty of 6 and 9 per cent respectively. In particular, the velocity dispersion, $\sigma_0$, has an uncertainty of 20 per cent. The major contribution to this error is the difference between the maximum a posteriori parameter value of 51.1 km s$^{-1}$ obtained by {\sc MultiNest} and the corresponding value of 42.6 km s$^{-1}$ obtained by the non-linear optimizer (see Section \ref{sec:results}). However, given the channel width of 33.9 km s$^{-1}$ for this system, we believe the discrepancy not to be significant.

\subsection{Mock dataset M7} 
\label{sec:m7}

The derivation of the rotation curve for low-inclination galaxies is challenging for any kinematic-fitting algorithm. For example, for $i=40^\circ$ an error as small as $\pm5^\circ$ in the estimation of the inclination angle can lead to a significant underestimation/overestimation of the rotation velocity as large as 10 per cent. To test the reliability of our modelling technique when the background source is a low-inclination galaxy, we have created the mock data M7 with the same lens and kinematic parameters of M5, but with an inclination angle for the source of 40$^\circ$, instead of 68$^\circ$.\\
As described above, we first model the zeroth-moment map and then use the recovered lens model parameters to derive a 3D model of the source. The latter is then analysed with $^{\mathrm{3D}}$BAROLO to obtain initial guesses for the source geometrical parameters. For the mock data M7, this results in a value of $i=41.5^\circ$, in close agreement with the input value. Subsequently, by applying our 3D lens modelling analysis to the full lensed data cube, we derive an inclination angle of 40.2$^\circ$. The inferred lens and kinematic parameters have median relative uncertainties of 8 and 6 per cent respectively, differing by 1-$\sigma$ and 2-$\sigma$ from the input values. We can conclude, therefore, that the accurate reconstruction of the zeroth-moment map allows to obtain a reasonable initial estimate of the inclination and, as a consequence, the kinematic properties of the galaxy are correctly recovered.

\subsection{Mock dataset M8} 
\label{sec:m8}

\citet{barolo} have shown that for large inclinations, $i>75^{\circ}$, the inner points of the rotation curve can be underestimated and that this effect can be more significant for a flat rather than for a solid-body rotation curve. This is due to the fact that $^{\mathrm{3D}}$BAROLO works ring by ring. However, we note that for a value of the inclination angle $>75^{\circ}$ the errors on the inclination have little impact on the derived rotation curve, due to the sinusoidal dependence between the line-of-sight velocity and the inclination, see equation (\ref{eq:vlos}). \\
These mock data were built to study how an extreme value of the inclination angle affects the reconstruction of the source kinematics. For this reason, we have built the mock data M8 using the same lens and kinematic models as those used for M2, but assuming an inclination angle of $80^{\circ}$. In particular, we focus on M2 because it has a flat rotation curve.\\
The inferred values that describe the rotation velocity differ by $\sim$4 per cent from the input values, reproducing very well also the inner regions of the curve (see the red dashed line in Figure \ref{fig:vel}). We can conclude, therefore, that the use of functional forms for the rotation velocity allows us to reproduce the inner regions of an edge-on galaxy better than a ring by ring method. Moreover, the inferred lens and kinematic parameters are characterised by a median relative uncertainties of 3 and 7 per cent and they are within 2-$\sigma$ from the input values.\\

\subsection{Mock dataset M9} 
\label{sec:m9}

\begin{figure}
  \includegraphics[width=0.92\columnwidth]{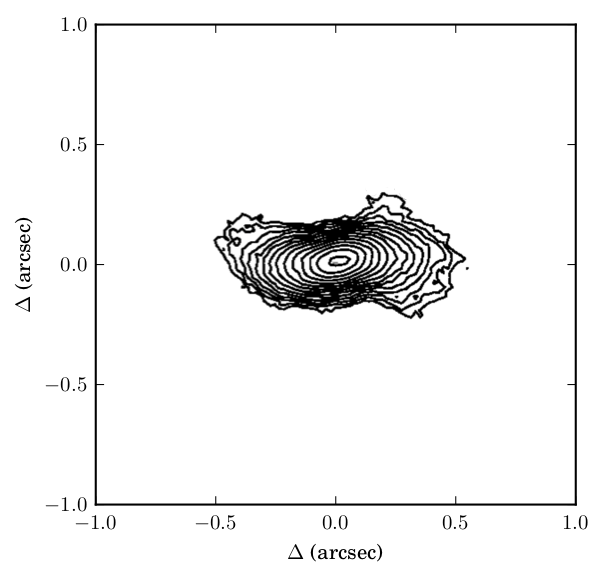}
  \caption{Zeroth-moment map for the reconstructed source M9, resulting after the first step (i.e. point \ref{item:1}, Section \ref{sec:optimisation}) of our optimisation strategy.}
  \label{fig:warp}
\end{figure}

The input lens and kinematic models are the same as those used to create the simulated data M5, but the input geometry of the source is different. In particular, the position angle has a strong warp, and it decreases linearly from a value of 280$^\circ$ at $R=0$ kpc to $250^\circ$ at the outermost radius.\\
Interestingly, we find that the presence of the warp is already revealed at the first step of our optimisation strategy (Figure \ref{fig:warp}), where we focus on the lens modelling of the zeroth moment (i.e. point \ref{item:1}, Section \ref{sec:optimisation}). From a $^{\mathrm{3D}}$BAROLO analysis of the derived source cube, we then obtain a position angle that changes linearly with a slope of 2.6 $^\circ$ kpc$^{-1}$ from an inner value of 278$^\circ$. We then apply our 3D lens modelling technique with a position angle which changes linearly. The two parameters that describe this change are left free to vary, starting from the initial guesses found by $^{\mathrm{3D}}$BAROLO. The best-fit slope that describes the variation of the $PA$ is 3.3 $^\circ$ kpc$^{-1}$,  which differs by $\sim$6 per cent from the input value of 3.5 $^\circ$ kpc$^{-1}$. The inferred value of $PA$ at $R=0$ kpc is 277.8$^\circ$, differing by $\sim$1 per cent from the input value of 280.0$^\circ$. The inferred kinematic parameters have a median uncertainty of 7 per cent, while the lens parameters have a median uncertainty of 6 per cent. Both the lens and the kinematic parameters are within 1-$\sigma$ from the input values.

\subsection{Further tests}
\label{sec:further_tests}

Observational evidence seems to indicate that physical processes such as disc turbulence, gas accretion and subsequent disc instabilities are more prevalent at high-redshift \citep[e.g.][]{law09, forster06, wisnioski}. As a consequence, the contribution of non-rotational components could have a significant impact on the kinematic of high-redshift galaxies.
The presence of significant radial motions, which are not included in the current analysis, could in principle limit the applicability of our technique to high-z galaxies. To quantify this issue, we have created a simulated dataset which has the same lens and kinematic models as M1 but includes radial motions of 25 kms$^{-1}$. We note that this value is larger than what is typically observed at low redshift \citep[i.e. $\lesssim10$ kms$^{-1}$,][]{tracht}, while, to date, there are no studies of radial motion in high-redshift galaxies. We have compared the lensed images of these simulated data with those of M1 for different values of angular resolution. Even in the ideal case of no observational noise, we have found no significant discrepancy, with a relative difference between the two lensed images of the order of $\sim$2 per cent, for the highest angular resolution case. We can conclude, that, although non-circular motions could contribute to the overall kinematics, they are mostly not detectable at the current angular and spectral resolution.

\begin{figure*}
  \includegraphics[width=2\columnwidth]{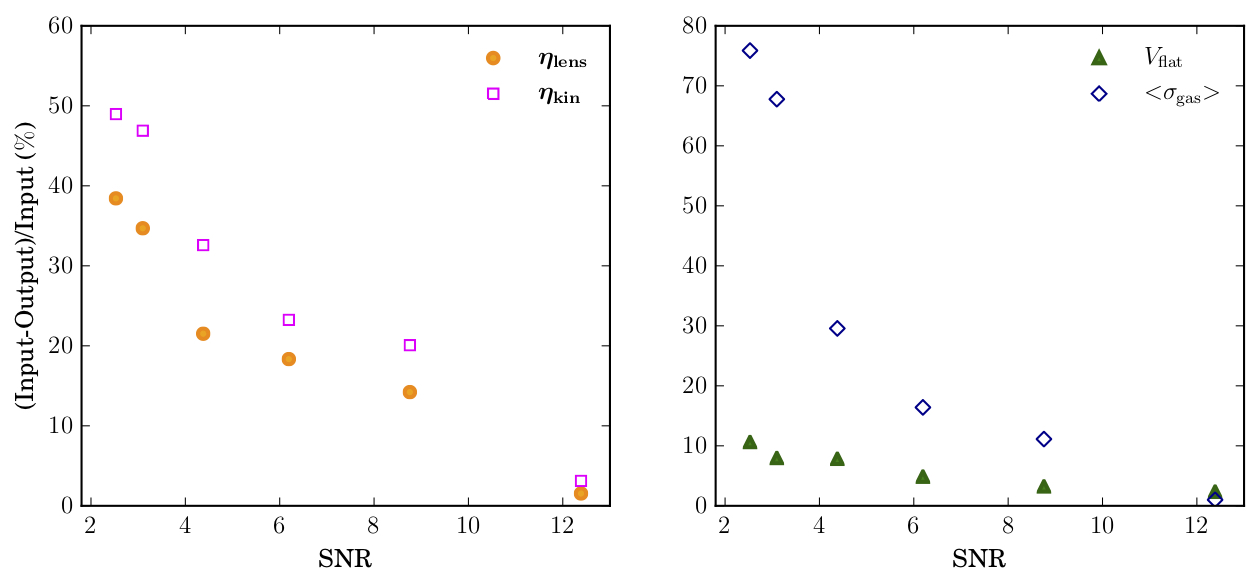}
  \caption{\textit{Left}: Relative difference between the input and the recovered lens (orange circles) and kinematic (magenta empty squares) parameters as a function of the signal-to-noise ratio for different data-quality realization of M1 mock dateset. \textit{Right}: same as in the left panel but for $V_{\mathrm{flat}}$ (green triangles) and $\mean{\sigma_{\mathrm{gas}}}$ (blue empty diamonds).}
  \label{fig:snr}
\end{figure*}

To test how the SNR of the data affects the accuracy with which the lens-mass and source-kinematics parameters are recovered, we have re-simulated the mock data M1 with five different noise levels, obtained using different exposure times (see Figure \ref{fig:snrm1} in appendix \ref{app:snr}). As shown in the left panel of Figure \ref{fig:snr}, for SNR\footnote{The values of the SNR are calculated as median of the SNR${_\mathrm{i}}$ for each spectral channel $\mathrm{i}$ of the data cube. The latter is calculated as SNR$_{\mathrm{i}}=\sum_{\mathrm{k}}S_{\mathrm{k i}}/\sqrt{N_{\mathrm{k i}}^2}$, where the sum is over the pixels of the $\mathrm{i}$th channel, $S_{\mathrm{k i}}$ is the signal at $\mathrm{k}$th pixel, and $N_{\mathrm{k i}}$ is its noise.} below $\sim$3 the relative difference between the input and the recovered values is higher than 30 per cent both for the lens (orange circles) and for the kinematic parameters (magenta empty squares). If we focus on the relative difference between the input and the output $V_{\mathrm{flat}}$ (green triangles on the right panel of Figure \ref{fig:snr}), we conclude that we are still able to recover it with an accuracy of the order of 90 per cent even for the mock dataset with the lowest SNR. The relative differences between the input and the recovered values of $\mean{\sigma_{\mathrm{gas}}}$ (blue empty diamonds on the right panel of Figure \ref{fig:snr}) are smaller than $\sim30$ per cent for SNR larger than 3.

So far, we have modelled all the simulated datasets with a fixed systemic velocity of zero km s$^{-1}$ and the centre of the kinematic prior set at the flux-weighted average position of the zeroth-moment map. However, for high-redshift galaxies, one expects error on the centre of the order of 5 to 10 per cent and on the systemic velocity of the order of the channel width. We have, therefore, repeated the analysis of M1 by fixing the coordinates of the centre to values which are offset from the real ones by 10 per cent and with values of the systemic velocity which are offset by the channel width velocity from the real value. We have found that an incorrect choice of the systemic velocity does not affect our results significantly, while changes in the centre have a significant effect only if both coordinates are shifted by 10 per cent relative to the real values in the same direction.

\begin{figure}
  \includegraphics[width=0.92\columnwidth]{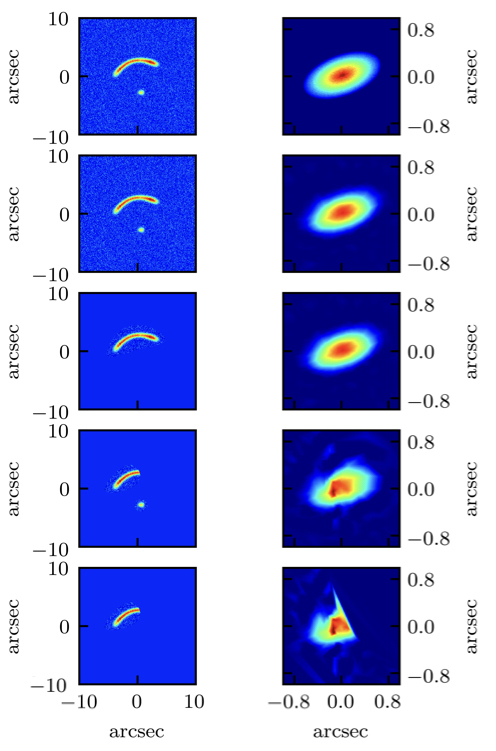}
  \caption{First row: on the left side the input lensed data obtained by lensing forward the disc galaxy on the right side. From second to fifth row: reconstruction of the source (right panels) obtained by modelling different parts of the mock lensed data (left panels). Second row: all data are considered for the source reconstruction. Third row: the counter image is excluded. Fourth row: the counter image and only part of the arc are modelled. Fifth row: only part of the arc is modelled to reconstruct the source.}
  \label{fig:test}
\end{figure}

Because of the small FOV of many IFU instruments, complete imaging of the lensed emission is often not possible, instead only a limited part of the arc is observed. Here, we show how this observational limitation can strongly affect the reconstruction of the source morphology. The first row of Figure \ref{fig:test} shows the emission of a background disc galaxy as lensed by a power-law mass distribution. In the same figure, we present the source that is reconstructed by considering an increasingly smaller part of the data. In particular, on the second row, we have modelled the entire set of images, on the third one we have excluded the counter image, on the fourth we have modelled only a small region of the main arc and the counter image, while on the fifth one only a small region of the main arc was taken into account. We find that one can safely ignore the counter image only if all images are modelled (see the third row of Figure \ref{fig:test}). However, failure to observe the entire main arc leads to an incorrect source reconstruction and a wrong estimate of both the centre of the galaxy and its extension, strongly affecting the derived kinematics (see the fifth row of Figure \ref{fig:test}). In this case, including the counter image is fundamental for a better reconstruction of the source (see the fourth row in Figure \ref{fig:test}). This result is due to the extension of the source, such that different regions of the galaxy are lensed into different regions and/or number of the images, depending on the lens system configuration. We conclude, therefore, that the analysis of the full set of lensed images (obtainable for example with mosaic observations) is crucial for a reliable derivation of the source properties.

\section{Summary and conclusion}
\label{sec:conclusions}
In this paper, we have presented a new method to model the kinematics of strong gravitationally lensed galaxies from 3D emission line observations. The technique is entirely Bayesian: a Bayesian penalty function allows us to simultaneously infer the lens-mass and the source-kinematics properties from the same 3D data cube, while the Bayesian marginalised evidence enables us to rank and compare different lens and kinematic models. This new approach is also grid-based and hierarchical. The source is reconstructed on a magnification-dependent Delaunay tessellation, and its kinematics represent the hyper-parameter of the source prior. The primary focus of this method is to provide a robust approach for studying the resolved kinematic properties of high-redshift lensed galaxies. In this respect, it represents a significant improvement over past and recent approaches. Indeed, since the kinematic model is a hyper-parameter of the reconstructed source, the inferred kinematic properties are not influenced by the poor understanding of the errors and spatial resolution on the unlensed plane. Furthermore, the lens mass distribution is derived consistently from the same 3D data cube.

To test the capabilities of this new method in inferring the correct model parameters, we have studied a sample of nine simulated lensed galaxies as they would be observed with the OSIRIS spectrograph. These galaxies are characterised by a variety of rotation curves and geometries. In particular, we have focused on rotation curves that are described either by different functional forms (i.e. simulated data M1-M3) or derived from real galaxies (i.e. datasets M4-M6). We have found that this variety of shapes for the input rotation curves (from solid-body to flat rotation curves) is robustly recovered. In particular, the median relative accuracy on the inferred lens and kinematic parameters are at the level of $\sim$1 and $\sim$2 per cent, respectively. 

Focusing on the extreme cases of a low-inclination (40$^{\circ}$, M7) and edge-on (80$^{\circ}$, M8) galaxy, we have also studied how the inclination of the source affects the accuracy of the reconstructions. We have found that the kinematic parameters can be recovered with a median accuracy of 1 and 2 per cent, respectively, if a reasonable initial estimate of the inclination can be obtained from the analysis of the zeroth-moment map. We have then tested the capability of our code to identify the presence of a warp. We have concluded that warps as large as 30 degrees can significantly affect the lensed data. However, we are still able to recover the kinematic parameters with an accuracy of 3 per cent.
The rotation curves in all cases are accurately reconstructed, therefore the most important physical parameters of the source galaxy (e.g., the dynamical mass, the angular momentum) can be correctly inferred.\\
We have also investigated the effect of increasing noise and concluded that the parameters are recovered with an accuracy better than 30 per cent
whenever the SNR is higher than $\sim$3. The flat part of the rotation velocity is recovered with an accuracy of the order of 90 per cent even when the SNR is $\sim 3$. Finally, we have examined the effect of strong radial motions and found it to be irrelevant for the typical angular and spectral resolution of IFU observations. From these extensive tests, we can conclude that the method presented in this paper offers a novel and robust way to study the gas kinematics of high-redshift lensed galaxies using data from the last generation of IFUs. Taking advantages of strong gravitational lensing we can study the kinematic properties of galaxies at z$\sim 2-3$ with spatial resolutions and SNR not achievable for unlensed galaxies, even with current observational technique. Moreover, gravitational lensing offers the unique opportunity to study galaxies in the low-stellar-mass range, which is almost impossible for studies targeting unlensed galaxies.

In this paper, we have focused on galaxy-scale lenses. The formalism of our method is applicable also to cluster lenses, although with a more complex parametrisation of the lensing potential. However, as the mass distribution of galaxy clusters is more complicated, we expect in this case larger uncertainties on both the lens parameters and the source kinematics.

We have also only focused on optical observations. However, our technique, being an extension to the 3D domain of the methods developed by \citet{vegetti09} and \citet{rybaka}, can also be used for the modelling of emission lines with radio interferometric observations. While our current implementation does not include functional forms for the analysis of radial motions and/or outflows, we plan to implement these in a follow-up work. In the near future, we will apply our novel technique to the analysis of both ionised and molecular gas emission lines to study the extended physical and kinematical properties of high-redshift lensed galaxies.\\
The code presented in this paper is not publicly available, however, the reader interested in using the code can contact the first author.

\section*{Acknowledgements}
We thank an anonymous referee for a very careful report and for his/her comments. We thank Cristopher Fassnacht and Tucker Jones for their helpful comments and suggestions on the mock data. SV has received funding from the European Research Council (ERC) under the European Union's Horizon 2020 research and innovation programme (grant agreement No 758853). EDT acknowledges the support of the Australian Research Council through grant DP160100723.



\bibliographystyle{mnras}
\bibliography{reference} 

\appendix
\section{Signal-to-noise ratio}
\label{app:snr}
\begin{figure*}
\includegraphics[width=1.8\columnwidth]{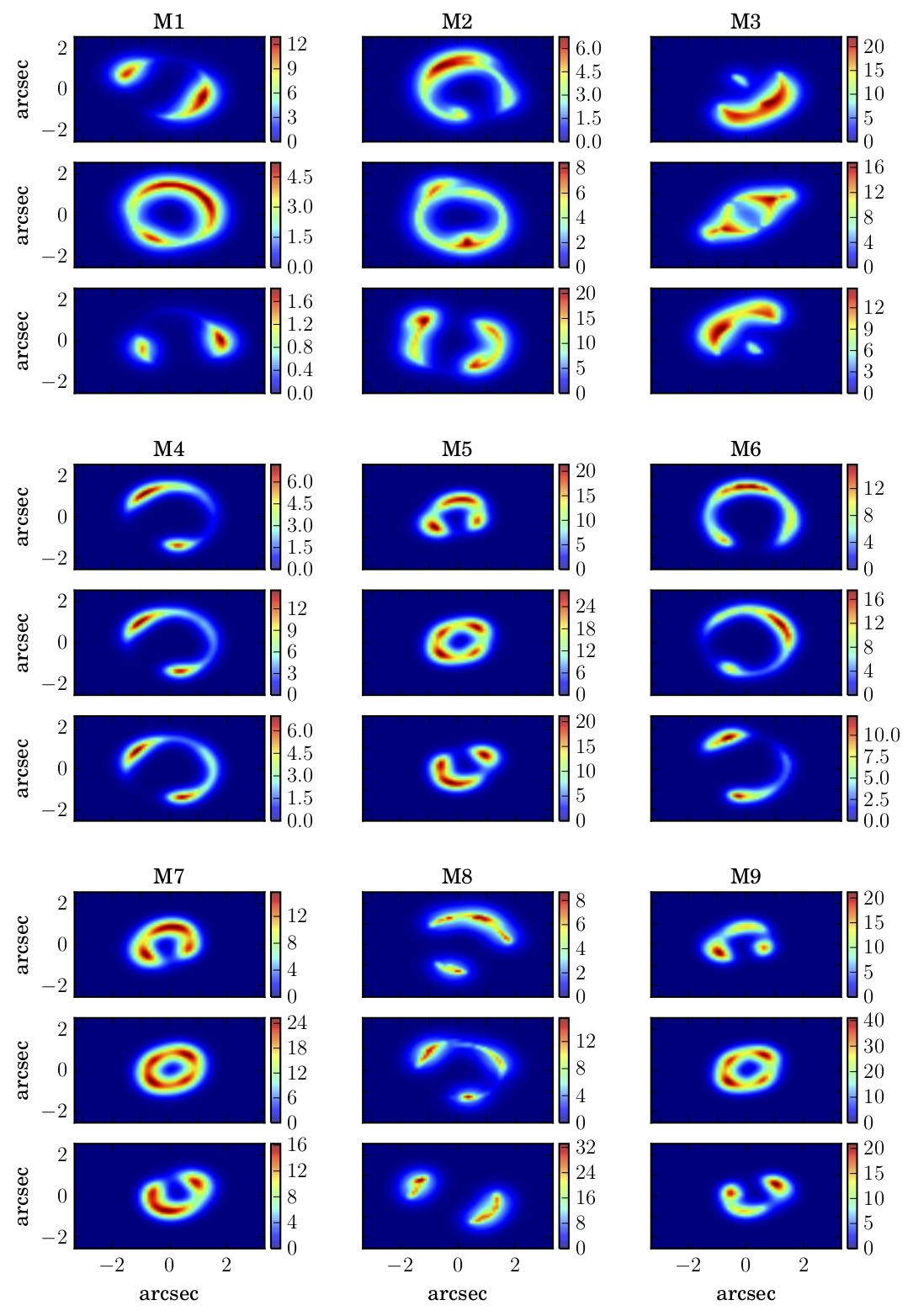}
\caption{SNR for M1-M9. The spectral channels are the same shown in Figure \ref{fig:m1} for M1 and \ref{fig:m2}-\ref{fig:m9} for M2-M9.}
\label{fig:snrm1m9}
\end{figure*}

\begin{figure*}
\includegraphics[width=1.5\columnwidth]{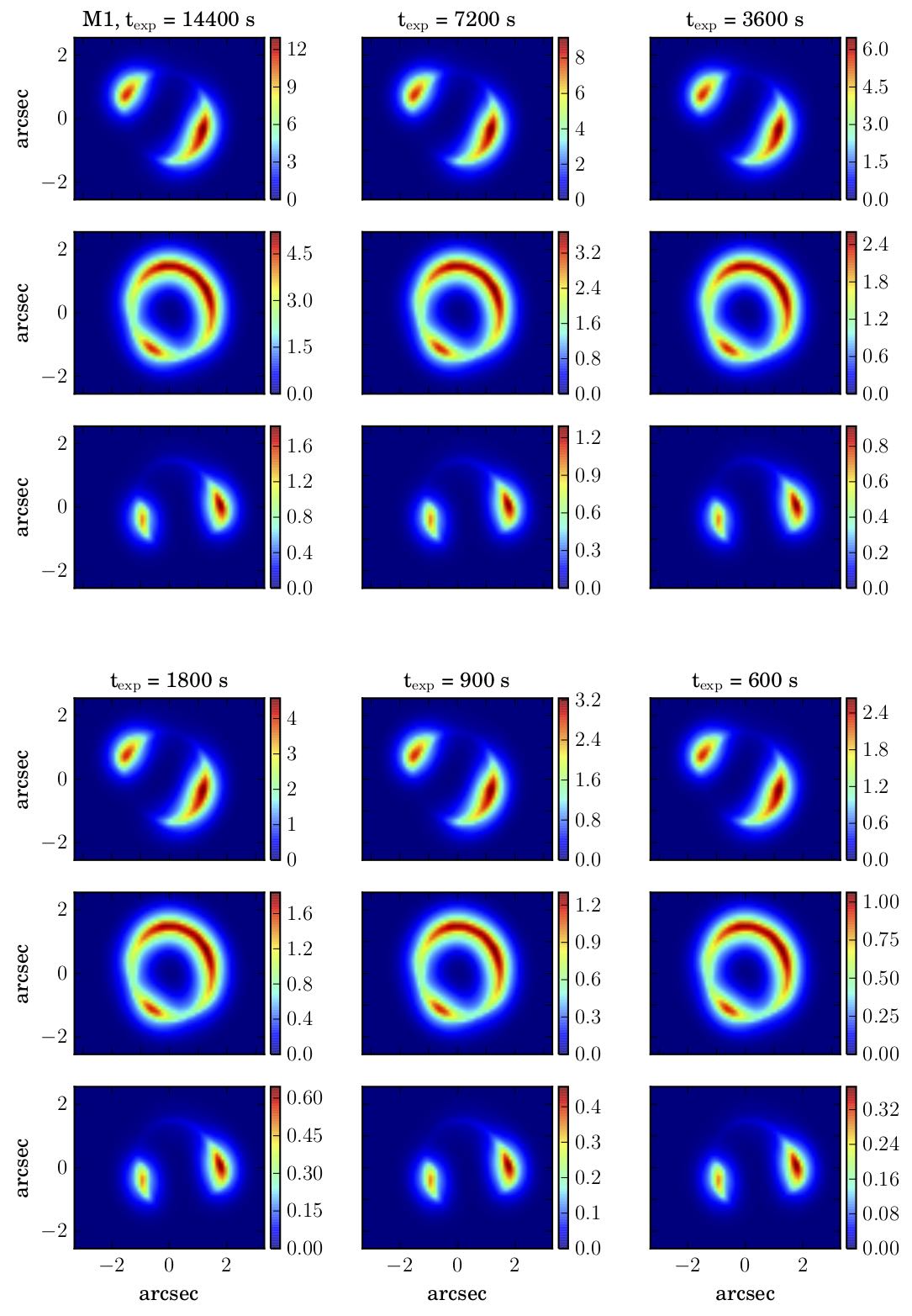}
\caption{SNR for the data-quality tests described in Section \ref{sec:further_tests} and in Figure \ref{fig:snr}. The different qualities of the data are obtained by using different exposure times, as indicated in the titles. The spectral channels are the same as shown in Figure \ref{fig:m1}.}
\label{fig:snrm1}
\end{figure*}

In Figure \ref{fig:snrm1m9} we show the SNR for the same spectral channels shown in Figures \ref{fig:m1} and \ref{fig:m2}-\ref{fig:m9}. For each spectral channels the noise is added considering the procedure described in Section \ref{sec:simulated} and the exposure time listed in Table \ref{tab:filter}.\\
In Figure \ref{fig:snrm1} we show the SNR for the same spectral channels shown in Figures \ref{fig:m1} and for the different values of data quality, as described in Section \ref{sec:further_tests} and summarised in Figure \ref{fig:snr}. The six different noise levels are obtained by considering six values of the exposure times for M1: 14.4 ks, 7.2 ks, 3.6 ks, 1.8 ks, 900 s and 600 s.

\section{Mock dataset M2-M9}
\begin{figure*}
\includegraphics[width=2\columnwidth]{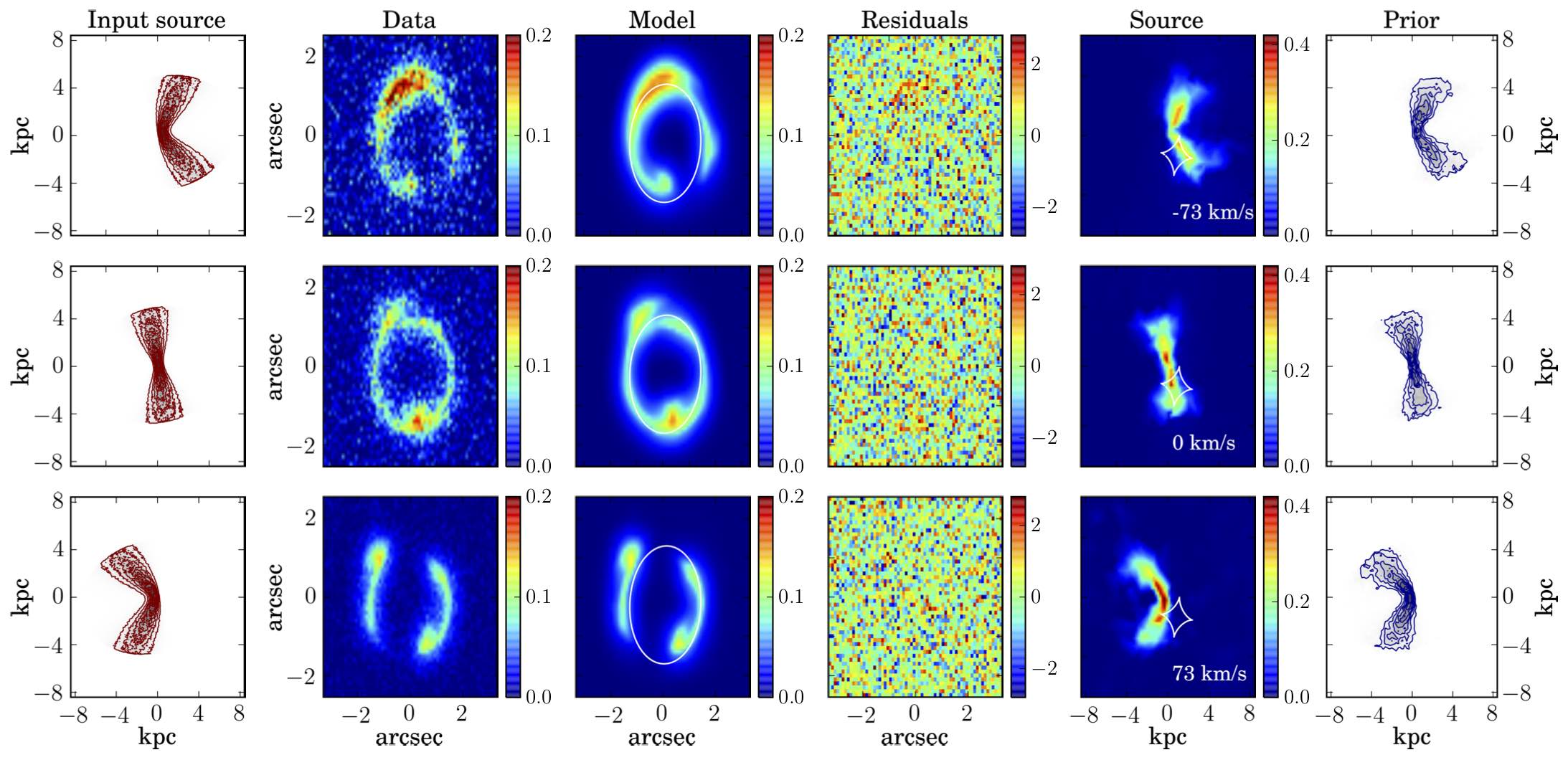}
\caption{Same as Figure \ref{fig:m1} for the simulated dataset M2, with $n=\left\{0.1, 0.2, 0.3, 0.4, 0.5, 0.6, 0.7, 0.8\right\}$.}
\label{fig:m2}
\end{figure*}

\begin{figure*}
  \includegraphics[width=2\columnwidth]{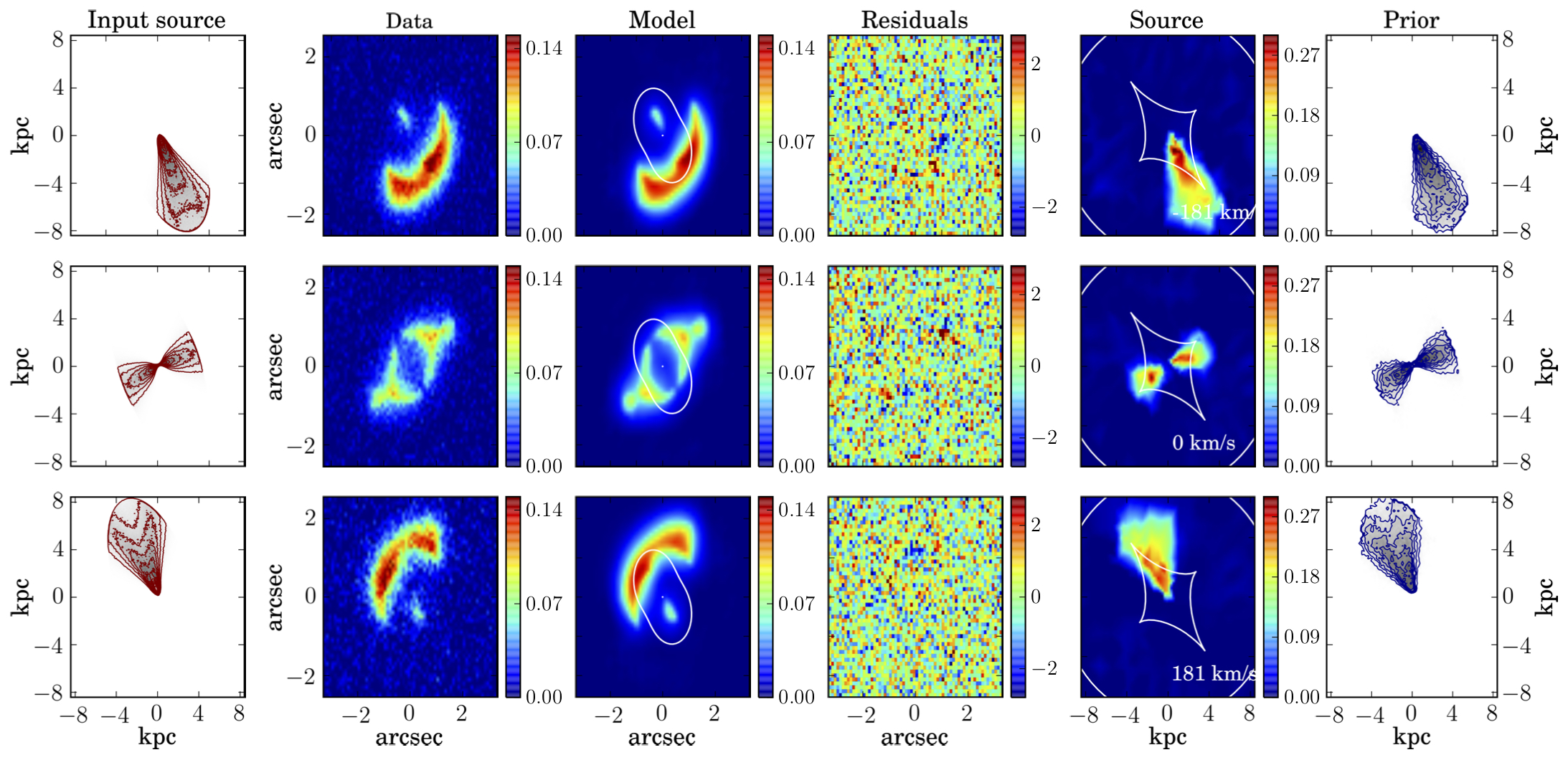}
  \caption{M3: same as Figure \ref{fig:m1} for the simulated dataset M3, with $n =\left\{0.1, 0.2, 0.4, 0.6, 0.8\right\}$.}
  \label{fig:m3}
\end{figure*}

\begin{figure*}
\includegraphics[width=2\columnwidth]{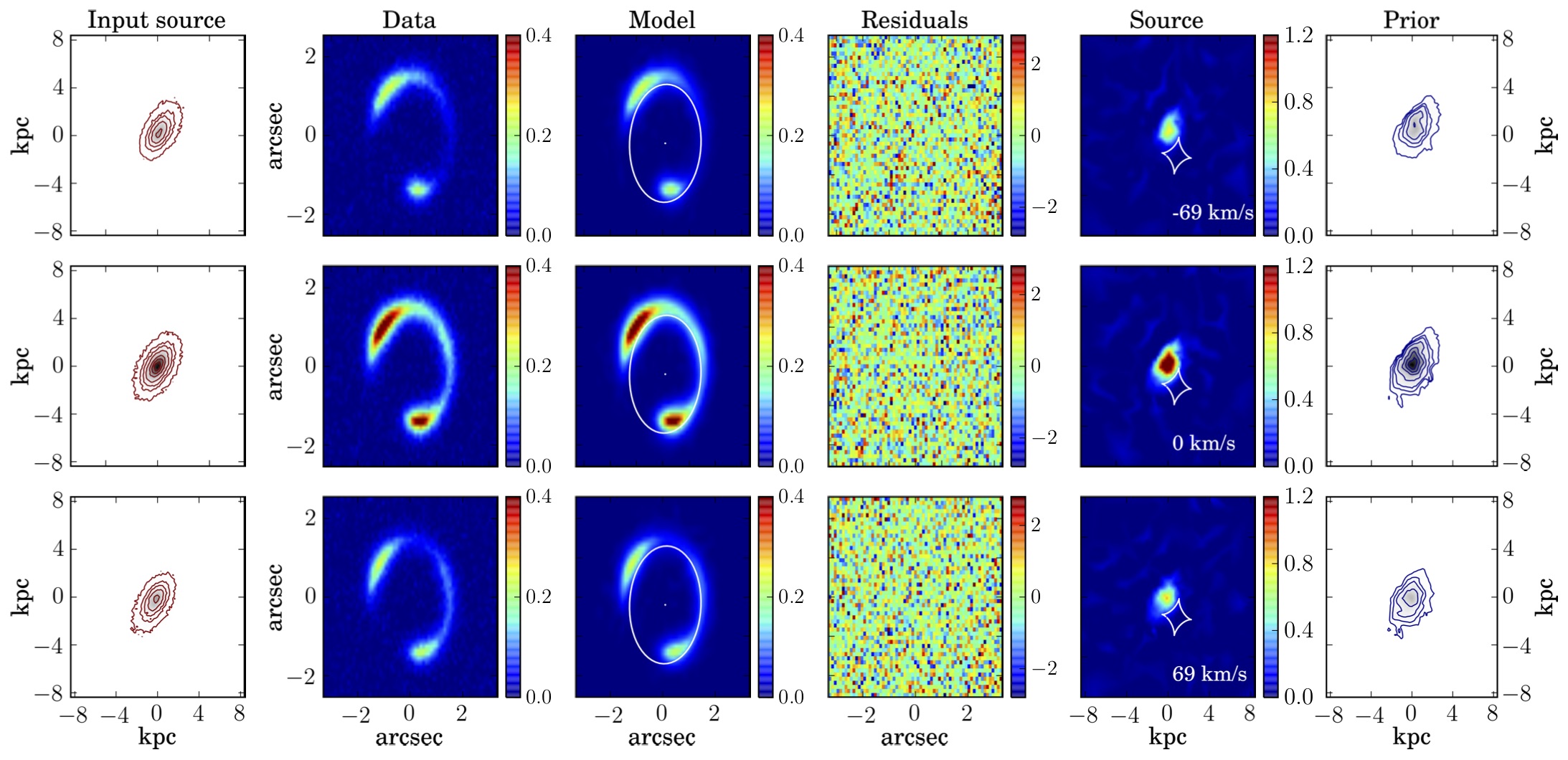}
\caption{Same as Figure \ref{fig:m1} for the simulated dataset M4, with $n=\left\{0.01, 0.05, 0.1, 0.2, 0.4, 0.6, 0.8\right\}$.}
\label{fig:m4}
\end{figure*}

\begin{figure*}
\includegraphics[width=2\columnwidth]{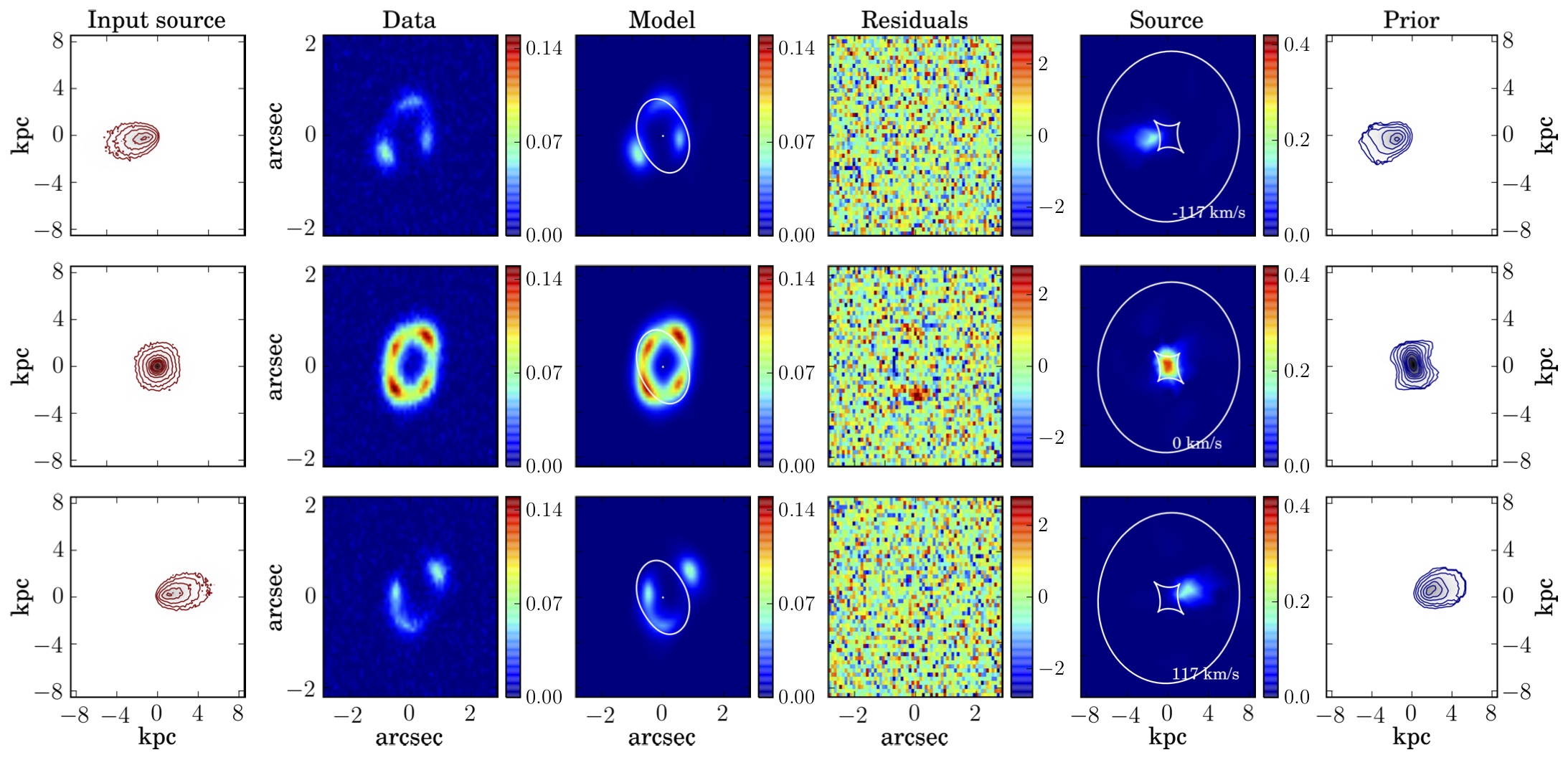}
\caption{Same as Figure \ref{fig:m1} for the simulated dataset M5, with $n=\left\{0.025 0.05, 0.1, 0.2, 0.3, 0.4, 0.6, 0.8\right\}$.}
\label{fig:m5}
\end{figure*}

\begin{figure*}
\includegraphics[width=2\columnwidth]{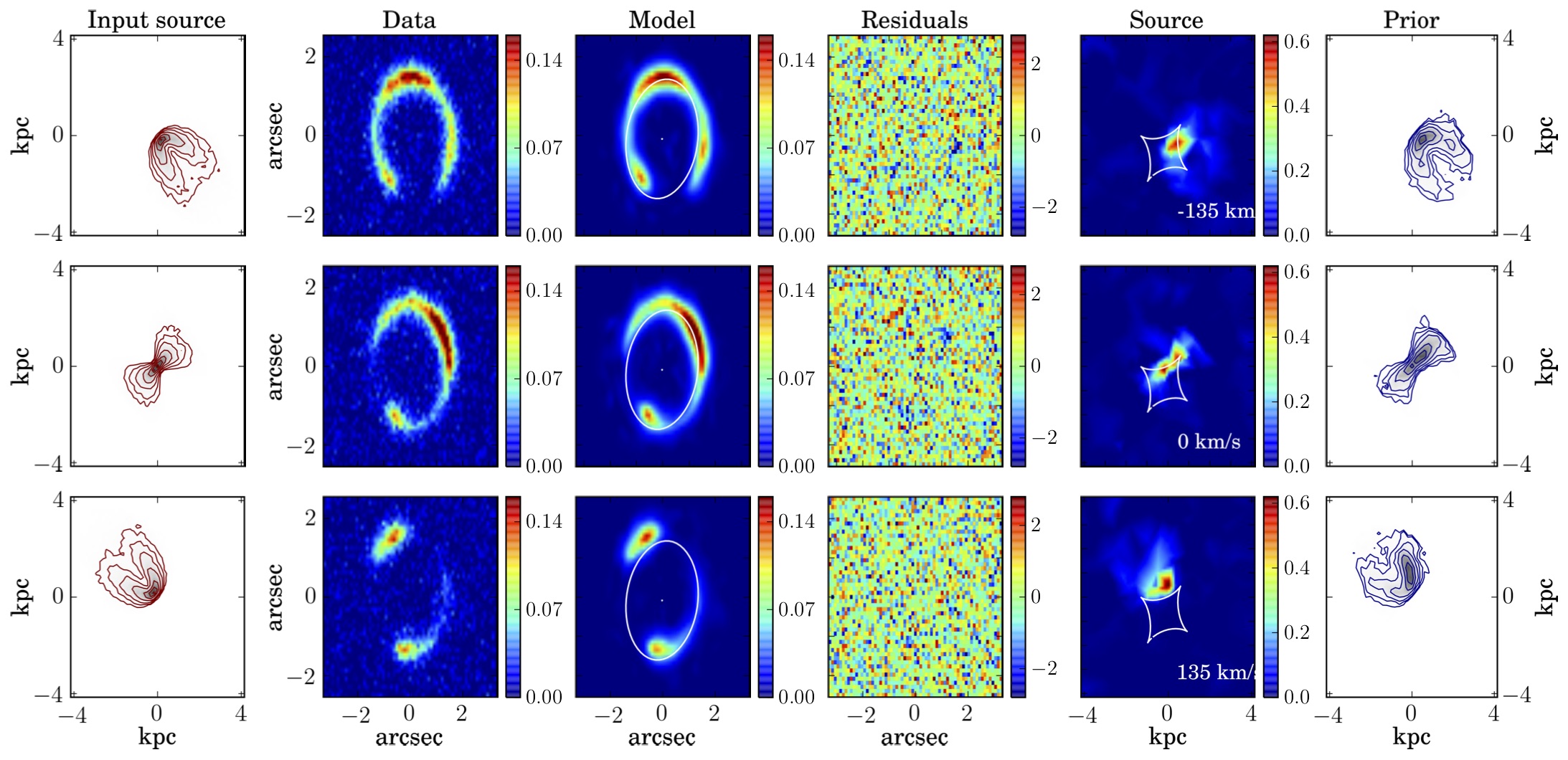}
\caption{Same as Figure \ref{fig:m1} for the simulated dataset M6, with $n=\left\{0.025, 0.05, 0.1, 0.2, 0.4, 0.6, 0.8\right\}$.}
\label{fig:m6}
\end{figure*}

\begin{figure*}
  \includegraphics[width=2\columnwidth]{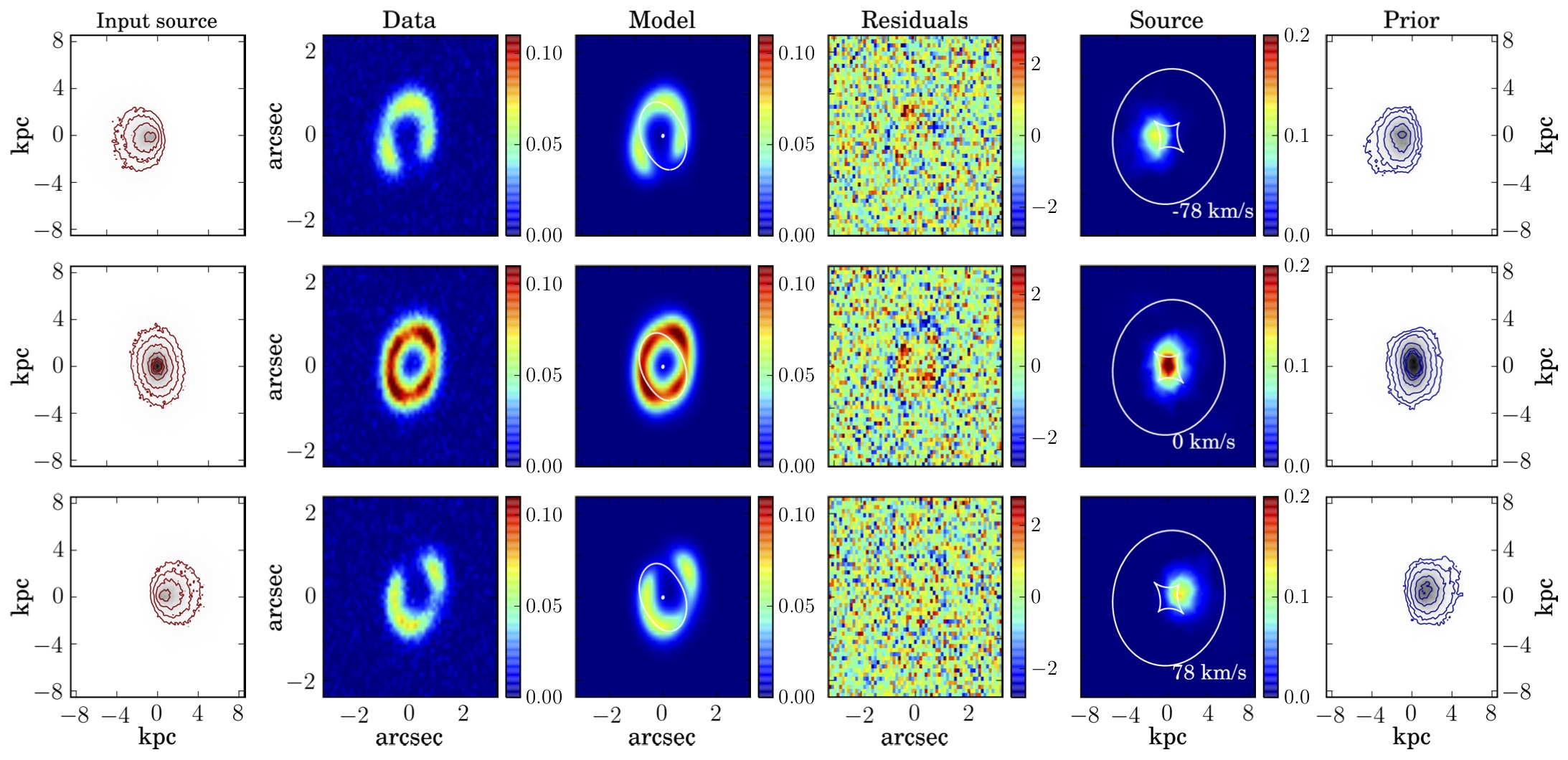}
  \caption{Same as Figure \ref{fig:m1} for the simulated dataset M7, with $n=\left\{0.09, 0.18, 0.36, 0.54, 0.64, 0.72\right\}$.}
  \label{fig:m7}
\end{figure*}

\begin{figure*}
  \includegraphics[width=2\columnwidth]{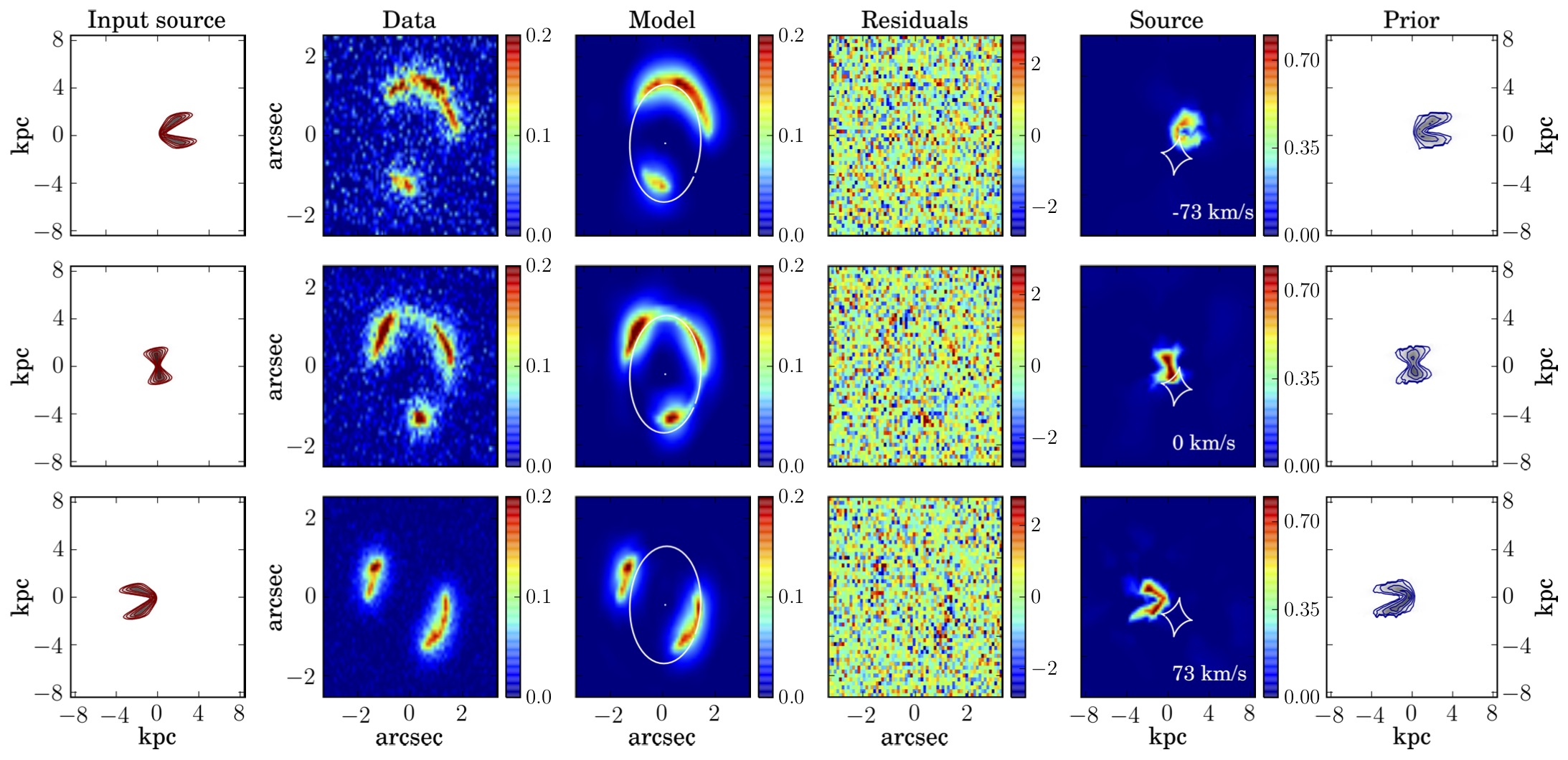}
  \caption{Same as Figure \ref{fig:m1} for the simulated dataset M8, with $n=\left\{0.1, 0.2, 0.4, 0.6, 0.8\right\}$.}
  \label{fig:m8}
\end{figure*}

\begin{figure*}
  \includegraphics[width=2\columnwidth]{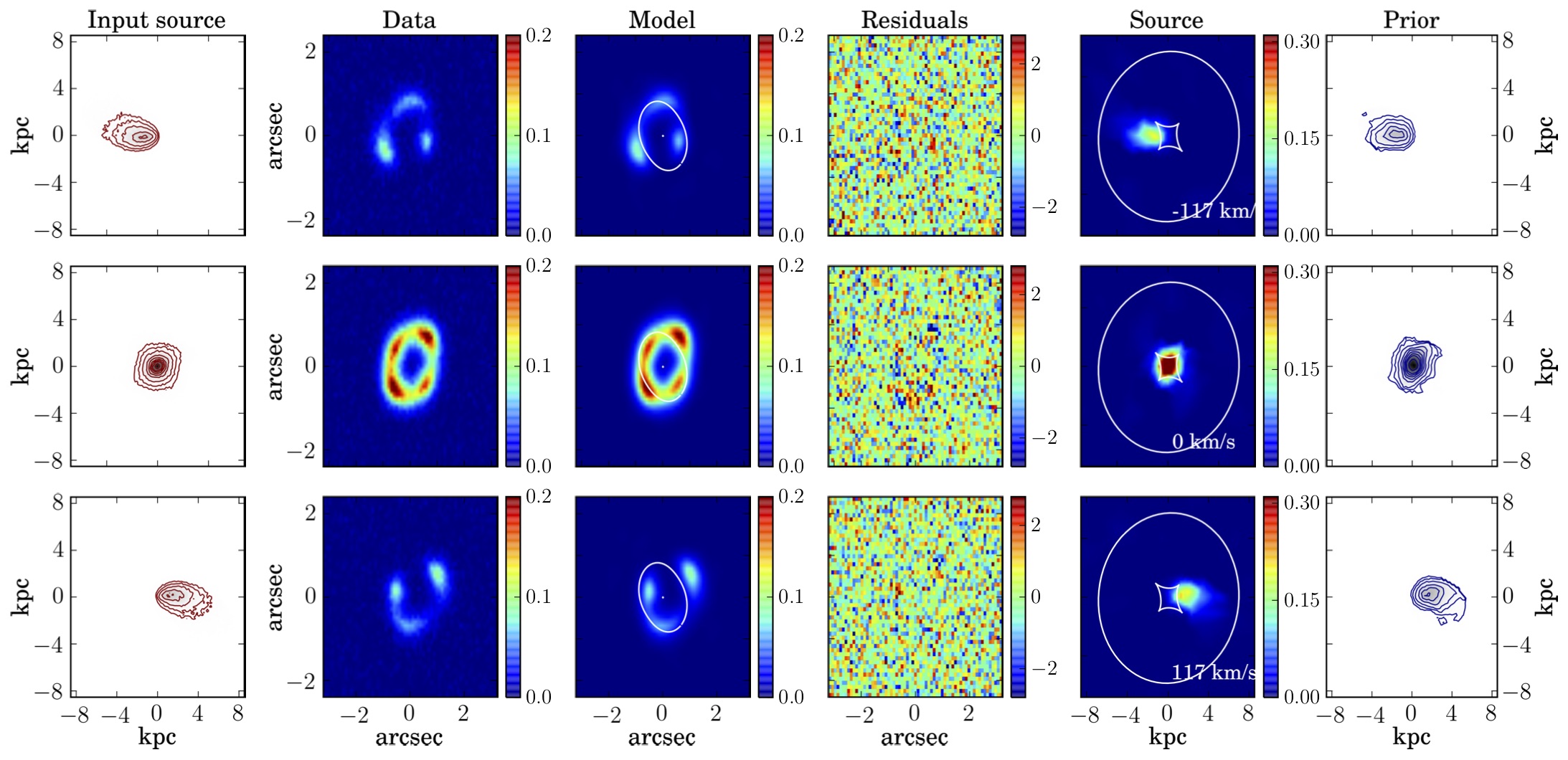}
  \caption{Same as Figure \ref{fig:m1} for the simulated dataset M9, with $n=\left\{0.1, 0.2, 0.4, 0.6, 0.8\right\}$.}
  \label{fig:m9}
\end{figure*}

As for Figure \ref{fig:m1} (for M1) we show in Figures \ref{fig:m2}-\ref{fig:m9} the input and best-fit models for M2-M8. For a selected number of spectral channels, we plot in the first column the contour levels of the input source, in the second column the simulated lensed data, in the third column the inferred lensed model, in the fourth column the normalised image residuals, in the fifth column the reconstructed source and in the sixth column the contour levels of the kinematic model.
\label{sec:appa}

\bsp  
\label{lastpage}
\end{document}